# A New Gate for Optimal Fault Tolerant & Testable Reversible Sequential Circuit Design

*A*

***Dissertation***

*Submitted*

*in partial fulfillment*

*for the award of the Degree of*

***Master of Technology***

*in **Department of Computer Science & Engineering***

***(With Specialization in Computer Engineering)***

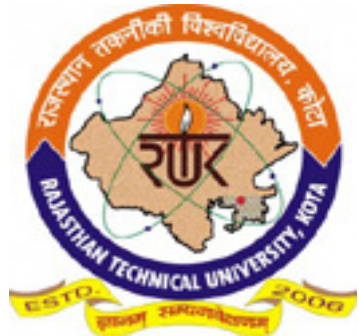

**Supervisor**

Dr. S.C. Jain

Professor, CSE

**Submitted By:**

Vishal Pareek

Enrollment No:

12E2UCCSM4XP619

**Department of Computer Science and Engineering**

**University College of Engineering**

Rajasthan Technical University

Kota (Rajasthan)

**SEPTEMBER 2014**

# Candidate's Declaration

I hereby declare that the work, which is being presented in the Dissertation, entitled "**A New Gate for Optimal Fault Tolerant & Testable Reversible Sequential Circuit Design**" in partial fulfillment of "Master of Technology" with specialization in Computer Science and Engineering, submitted to the Department of Computer Science & Engineering, University College of Engineering, Rajasthan Technical University, Kota is a record of my own investigations carried under the guidance of **Dr. S.C. Jain** Professor, Computer Science, University College of Engineering, RTU Kota. I have not submitted the matter presented in this Dissertation Report any where for the award of any other degree.

**Vishal Pareek**
Computer Science Engineering
Enrollment No.: 12E2UCCSM4XP619
University College of Engineering, RTU, Kota (Raj)

**Counter Signed by**
Supervisor

**Dr. S.C. Jain**
Professor,
Department of Computer Science & Engineering,
University College of Engineering,
Kota (Rajasthan)

# Certificate

This is to certify that this Dissertation entitled "**A New Gate for Optimal Fault Tolerant & Testable Reversible Sequential Circuit Design**" has been successfully carried out by Vishal Pareek (Enrollment No.: 12E2UCCSM4XP619), under my supervision and guidance, in partial fulfillment of the requirement for the award of Master of Technology Degree in Computer Science & Engineering from University College of Engineering, Rajasthan Technical University, Kota for the year 2012-2014.

**Dr. S.C. Jain**

Professor,

Department of Computer Science & Engineering,

University College of Engineering,

Kota (Rajasthan)

# Acknowledgments

It is matter of great pleasure for me to submit this report on dissertation entitled "**A New Gate for Optimal Fault Tolerant & Testable Reversible Sequential Circuit Design**", as a part of curriculum for award of "Master in Technology with specialization in Computer Science & Engineering" degree of Rajasthan Technical University, Kota.

I am thankful to my Dissertation guide **Dr. S.C. Jain**, Professor in Department of Computer Science for his constant encouragement, able guidance and for giving me a new platform to build by career by giving me a chance to learn different fields of this technology. I am also thankful to the Head of Computer Science Department for their valuable support.

I would like to acknowledge my thanks to entire faculty and supporting staff of Computer Engineering Department in general and particularly for their help, directly or indirectly during my Dissertation work.

I express my deep sense of reverence to my parents, family members and my friends for their unconditional support, patience and encouragement.

DATE                                                                                          Vishal Pareek

# List of Figures

# List of Tables

# CONTENTS

# ABSTRACT

With phenomenal growth of high speed and complex computing applications, the design of low power and high speed logic circuits have created tremendous interest. Conventional computing devices are based on irreversible logic and further reduction in power consumption and/or increase in speed appears non-promising. Reversible computing has emerged as a solution looking to the power and speed requirements of future computing devices. In reversible computing logic gates used are such that input can be generated by reversing the operation from output. A number of reversible combinational circuits have been developed but the growth of sequential circuits was not significant due to feedback and fan-out was not allowed. However, allowing feedback in space, a very few sequential logic blocks i.e. flip-flops have been reported in literature. In order to develop sequential circuits, flip-flops are used in conventional circuits. Also good circuit design methods, optimized and fault tolerant designs are also needed to build large, complex and reliable circuits in conventional computing. Reversible flip-flops are the basic memory elements that will be the building block of memory for reversible computing and quantum computing devices. In this dissertation we plan to address above issues. First we have proposed a Pareek gate suitable for low-cost flip-flops design and then design methodology to develop flip-flops are illustrated. Further almost all flip-flops and some example circuit have been developed and finally these circuits have been converted into fault tolerant circuits by preserving their parity and designs of offline as well as online testable circuits have been proposed.

Reversible computing can be used to develop logic level design of circuits. For implementation purpose many technologies are under development. Quantum computing devices are potential candidate for implementation of reversible logic. Hence, for comparing reversible circuits quantum cost is the one of the important parameter. In this dissertation work, we have also compared quantum cost as well as other parameters with existing circuits and shown a significant improvement in almost all parameters.

# INTRODUCTION

The high speed computation and complex computing application requirements are growing phenomenally in current compute intensive world. Fast growing computing demands the power consumption, heat dissipation and chip size issues are posing challenges for logic design with conventional technologies [1]. So the designs of low power and high speed logic circuits are creating tremendous interest in current scenario. Reversible computing is emerging as an alternative that offers high computation speed, high packaging density, low heat dissipation and low power consumption etc. A number of combinational circuits have been developed but the growth of sequential circuits was not significant due to feedback and fan-out constraint in reversible scenario. Reversible flip-flops are the most significant and basic memory elements that will be the target building block of memory for the forthcoming nanoscale devices. In order to solve any problem through reversible computing, it is essential that sequential reversible building block developed and used properly. The logic circuit design which provides high packaging density may not remain free from faults. In reversible computing to achieve reliable circuit design it is necessary to incorporate fault tolerance in circuit design. This dissertation addresses the problem of designing optimal sequential circuits and fault tolerance issues.

This chapter starts with motivation based on historical development and issues in conventional computing in section 1.1 followed by problem statement in section 1.2. Section 1.3 introduces salient contributions of the thesis, and finally section 1.4 states the outline of the dissertation.

## 1.1 Motivation

The computing word has advanced in computing power, higher transistor densities within given quantity of time, space and cost. The availability of low cost computing technology enables growth of new applications in various fields thus driving up demand for high computing power. Hardware technology has changed from vacuum tubes and transistor to multi-million gate solid-state devices to cope with power and densities constraints. Due to power-density constraints, physical limits of conventional computing are likely to halt transistor shrinkage for all major circuit types and technologies by 2021 [2]. In 1960 Gordon

Moore [3] predicted that the number of transistor counts in a device will double in every 18 months on average. As the number of transistor counts in a device increases the power dissipation of the device also increases. ITRS [4] has shown a road map of minimum feature size (transistor counts) according to future need at atomic level in 2030 as shown in figure 1.1.

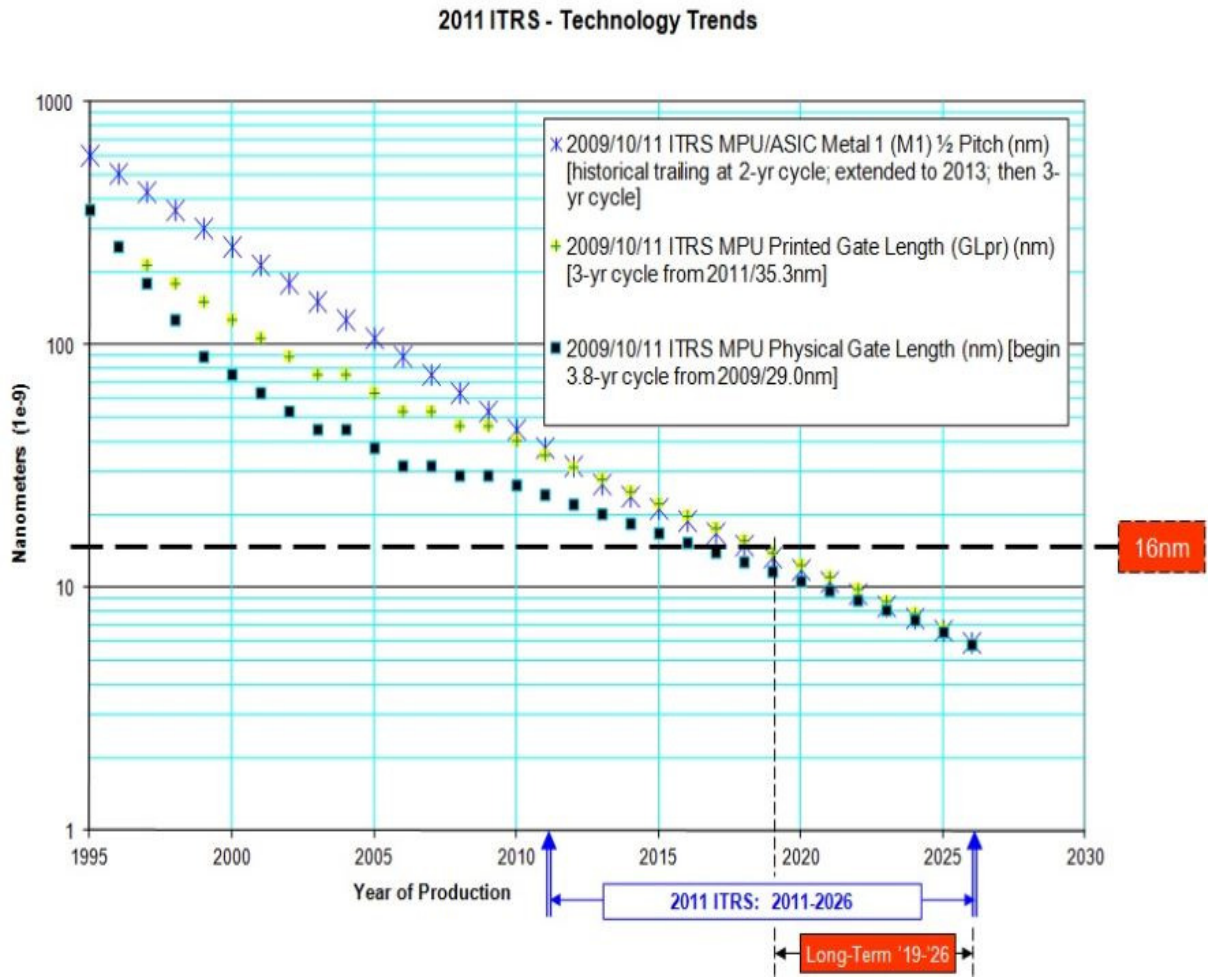

Figure 1.1   ITRS Feature Size Projection

As shown in Figure 1.2 ITRS has also shown a trend of minimum transistor switching energy for future need. The VLSI chip industry is moving at high pace towards miniaturization. With miniaturization it faces issue of large amount of heat dissipation. In the conventional technology, the logic blocks are normally irreversible in nature and according to Landauer's [5] computation with irreversible logic results in energy dissipation due to heat loss. Each bit of information dissipates at least $kT\ln 2$ Joules of energy where k is Boltzmann's constant and T is the absolute temperature at which the operation is performed. In early 1973, C. H. Bennett [6] had shown that the problem of heat dissipation of VLSI (Very large Scale

Integrated Circuits) can be overcome by using reversible logic because reversible computation does not require erasing any bit of information.

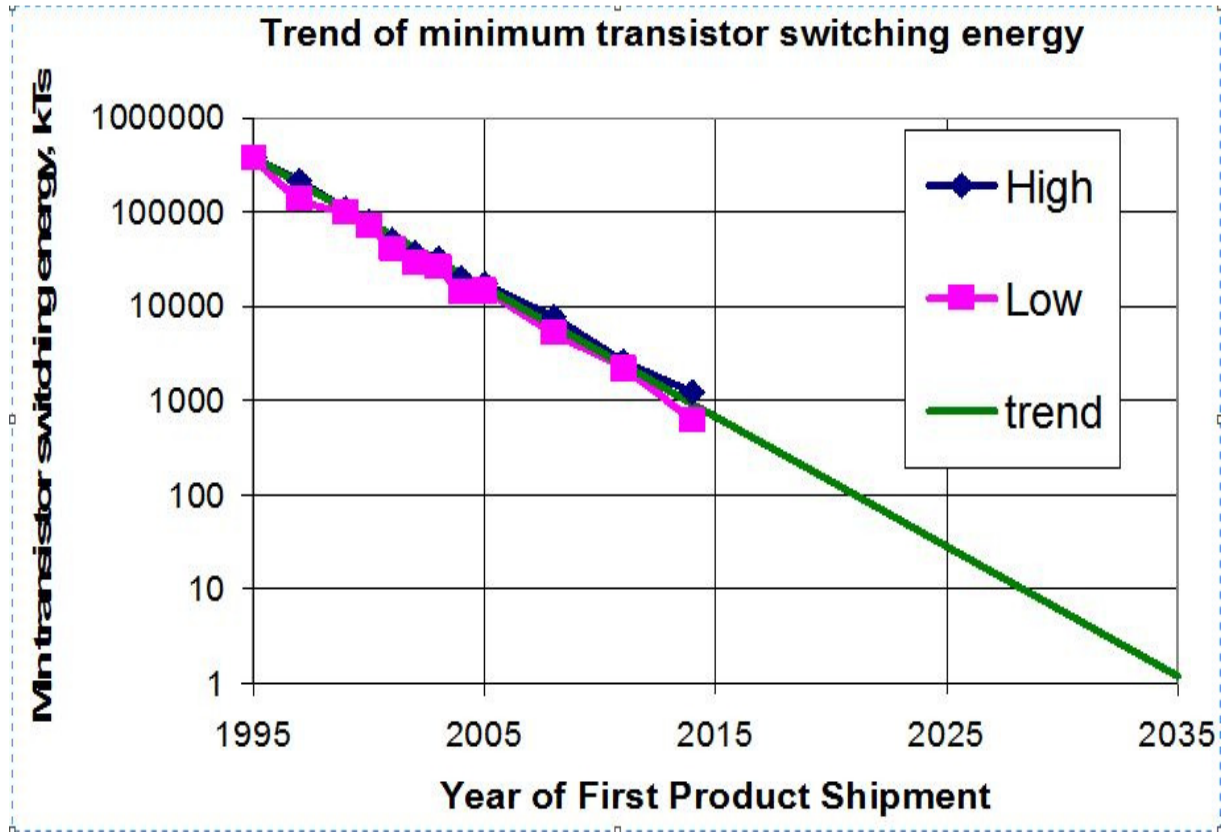

Figure 1.2   ITRS Trend of Minimum Transistor Switching Energy

Due to this fact, the loss of information and consequently dissipation of energy in computational operation is significantly lower than conventional logic. Hence, reversible computing is the best alternative for all future low power high speed technologies. This concept has worked as one of the motivations behind the present dissertation and many other related works. A number of combinational circuits have been developed but the growth of sequential circuits was not significant due to feedback and fan-out was not followed reversibility basics. A pioneering step in development of reversible sequential circuits was due to Toffoli [7] who had shown that reversible sequential circuits can be constructed provided the transition function of the circuit block without the feedback loop is unitary. Further it is proved that in order to make reversible finite automata one can build a reversible realization of its transition function and use it as a combinational part of the required sequential circuit. His ideas on the reversible sequential circuit had further strengthened in his opening up work on conservative logic [8]. Recently, A. Banerjee et al. [9] have redefined that feedback is allowed in space but not in time.  Hence, the development of reversible sequential circuits has begun. To synthesize reversible sequential circuits, reversible sequential basic building blocks such as latches and flip-flops are essential with optimized design parameters. Once a reversible sequential circuit is synthesized we require evaluating its quality. Different optimization parameters such as gate count (circuit cost), number of garbage output, number of constant input, quantum cost, hardware complexity are proposed. A considerable amount of work has been done to optimize a reversible circuit with respect to a particular optimization parameter. Once a circuit is realized and optimized it may be required to detect different kind of faults in that circuit [10]. This requirement has motivated behind the present thesis.

## 1.2 Problem Statement

It is proposed to realize optimized reversible sequential building blocks with fault tolerance capability to incorporate fault tolerance in reversible sequential circuit. The main work of this dissertation is to design reversible sequential building blocks with optimized parameters and incorporate fault tolerance in these elements by making them parity preserving and proposed offline and online testing designs.

## 1.3 Salient Contributions of the Dissertation

This dissertation makes significant contributions to the realization of reversible sequential building blocks, the synthesis of reversible sequential circuits, the design of fault tolerance reversible sequential building blocks, and testable design of reversible D flip-flop. The contributions to the realization of reversible sequential building blocks include:

- ➢ A novel parity preserving reversible gate which directly implement D flip-flop with minimum optimization parameters
- ➢ By using proposed reversible gate optimized design of reversible building blocks such as D flip-flop, R-S flip-flop, JK flip-flop and T flip-flop

This dissertation also presents synthesis of reversible sequential circuits which include:

- ➢ Optimized design of shift registers
- ➢ The first design of shift counter (Johnson counter)

The contributions to the design of fault tolerance reversible sequential blocks include:

> ➢ The first coherent designs of reversible sequential building blocks which incorporate fault tolerance by parity preserving characteristics.

This is a significant contribution, for instance, for the design of fault tolerance reversible sequential circuits which is necessary for high densities devices.

The final set of contributions of this work is to the design of testable reversible D flip-flop. These contributions include:

> ➢ Offline Testing for D flip-flop:  The design of reversible D flip-flop that can be tested by only two test vectors, for any stuck at faults.
> ➢ Online Testing for D flip-flop: The first design of reversible D flip-flop which online test single bit fault.

## 1.4 Outline of the Dissertation

This dissertation summarizes the previous work done reversible sequential circuits and presents realization of sequential circuit with fault tolerance capability. Chapter 2 gives foundation of reversible sequential logic, basic reversible gates, fault model in reversible computing and testing approaches for reversible sequential circuits. Previous work is also analyzed and summarized in this chapter as related work. As an important part of this summary, the weaknesses of previous work are pointed out. Chapter 3 contains the description and the analysis of the proposed design of reversible sequential elements. Chapter 4 describes the description of the proposed deign of fault tolerant and testable design of reversible sequential building blocks. Chapter 5 results, propose a comparative and statistical study of proposed realization of sequential building blocks with existing designs. In Chapter 6, the concluding chapter, we give some indication of the further possible work which could extend the results contained in this thesis.

# BACKGROUND & LITERATURE SURVEY

In this chapter, we first cover the fundamental concepts behind the reversible sequential logic. Then, we will discuss sequentially major aspects of reversible logic design in second section of this chapter with following main categories:

➢ Reversible Gate
➢ Basic Reversible Gate
➢ Reversible Circuit
➢ Optimization Parameters

Reversible computing has some fault models like conventional computing. In order to take advantage of these fault models in fault detection we require specialized mechanisms and techniques. Offline testing and online testing approaches available in literature for combination reversible logic will be described at last of this subsection. To cover these aspects, points are includes in third section of this chapter:

➢ Parity Preserving Reversible Logic Gates
➢ Conservative Reversible Logic Gates
➢ Parity Preserving Reversible Circuits
➢ Conservative Reversible Circuits
➢ Fault Models in Reversible Logic
➢ Fault Testing for Reversible Logic

The last section of this chapter provides a literature survey of previously proposed work in reversible sequential logic circuits. We categorize survey in the following categories:

➢ Reversible Sequential Building Blocks
➢ Reversible Sequential Shift Register &  Shift Counter
➢ Fault Tolerance Design of Sequential Building Blocks
➢ Survey Extraction

## 2.1  Reversible Sequential Logic

In the design of reversible circuits, the research work was primarily bounded to combinational circuit design, due to convention that feedback is not allowed in reversible logic. To be precise, conceptual issues related to sequential reversible logic feedback and fan-out will be discussed in following subsections.

### 2.1.1 Feedback in Reversible Sequential Logic

In one of the pioneer papers, Toffoli [7] has proved that feedback can be used in reversible logic. According to Toffoli, " A sequential circuit is reversible if its transition function is constructed by reversible logic." Further it is shown that to realize a reversible finite automaton one can construct a reversible realization of its transition function and use it as a combinational part of the required sequential circuit.

In 2010 Banerjee et al [9] has shown that feedback is not allowed in reversible logic if we consider feedback in a similar fashion as it is consider in conventional computing. Two objections against feedback have addressed. First, in a reversible circuit merging of two computational paths is not allowed and second, time axis goes from left to right in a reversible circuit. These objections against feedback circumvented by allowed feedback loop only in space but not in time.

### 2.1.2 Fan-out in Reversible Sequential Logic

According to fundamental of reversible logic fan-out structure is not reversible [11]. For fan-out, the number of input signal is one, but there are two or more output signals. Therefore, for the basic of reversibility, we need to make equal number of inputs and output signals. A Feynman Gate is used to this purpose to duplicate a signal.

## 2.2 Reversible Logic Design

The fundamental principle of reversible logic is that a bijective device with an identical number of input and output signals [7]. Fault tolerance reflects robustness in the system. Reversible gate, reversible circuit, optimization parameters, fault tolerance and testing approaches will be discussed in following subsections.

### 2.2.1 Reversible Gate

A Reversible Gate is a p-input, p-output (denoted by p × p) circuit that produces a unique output pattern for each possible input pattern [11]. There is a one to one correspondence between the input and output vectors. Different reversible gate under following categories are available in literature.

### 2.2.2 Basic Reversible Gates

In the literature, several 3×3 reversible gates such as the Toffoli gate [7], the Fredkin gate [8], and the Peres gate [12] have been addressed. Each reversible gate has a cost associated with it called the quantum cost. The quantum cost of a reversible logic gate is the count of 1×1 and 2×2 quantum logic gates or reversible logic gates required in its design. The quantum costs of all reversible 1×1 and 2×2 gates considered as unity. There are three (Controlled NOT gate (CNOT) commonly called Feynman gate, Controlled-V and Controlled-V⁺) basic 2×2 reversible gates. Any reversible gate can be realized using the NOT gate, and 2×2 reversible gates. Thus, in precise, the quantum cost of a reversible logic gate can be calculated by counting the numbers of NOT, Controlled-V, Controlled-V⁺ and Feynman gates required in its implementation.

➢ **The NOT Gate:** A NOT gate is a 1×1 gate represented as shown in Figure 2.1. It is a 1×1 gate, so its quantum cost is one.

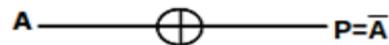

Figure 2.1   A NOT gate

➢ **Controlled-V and Controlled-V⁺ Gates:** The controlled-V gate is shown in Figure 2.2(a). In the controlled-V gate, when the control line A = 0 then the qubit B will pass, i.e., Q = B. When A = 1 then the unitary operation $V = \frac{i+1}{2}\begin{pmatrix} 1 & -i \\ -i & 1 \end{pmatrix}$ is operated to input B, that is, Q = V (B). The controlled -V⁺ gate is shown in Figure 2.2(b). In the controlled-V⁺ gate when the control line A = 0 then the qubit B will pass, means, Q = B. When A = 1 then the unitary operation $V^+ = V^{-1}$ is applied to the input B, that is, Q = V⁺ (B).

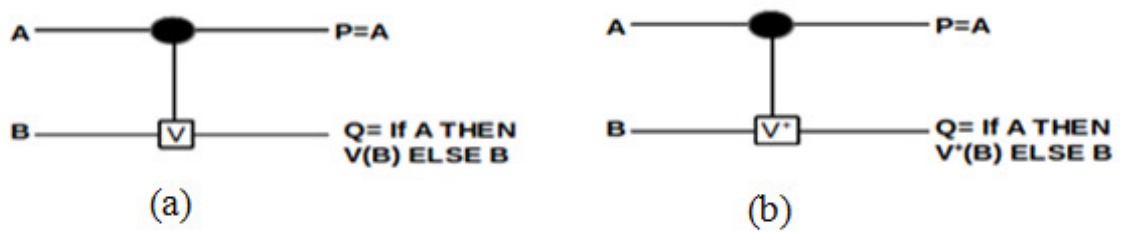

Figure 2.2   (a) Controlled - V (b) Controlled – V⁺

➢ **Feynman Gate (CNOT Gate):** The Feynman gate (FG) [13] or CNOT gate is a 2×2 reversible gate having the mapping (A, B) to (P = A, Q = A⊕B) where input signals are A, B and P, Q are the output signals, respectively. It is a 2×2 gate, so its quantum cost of unity. Figures 2.3(a), 2.3(b) and 2.3(c) show the truth table, block diagram and quantum implementation of the Feynman gate. The Feynman gate can be used for copying the signal thus avoiding the fan-out problem in reversible logic.

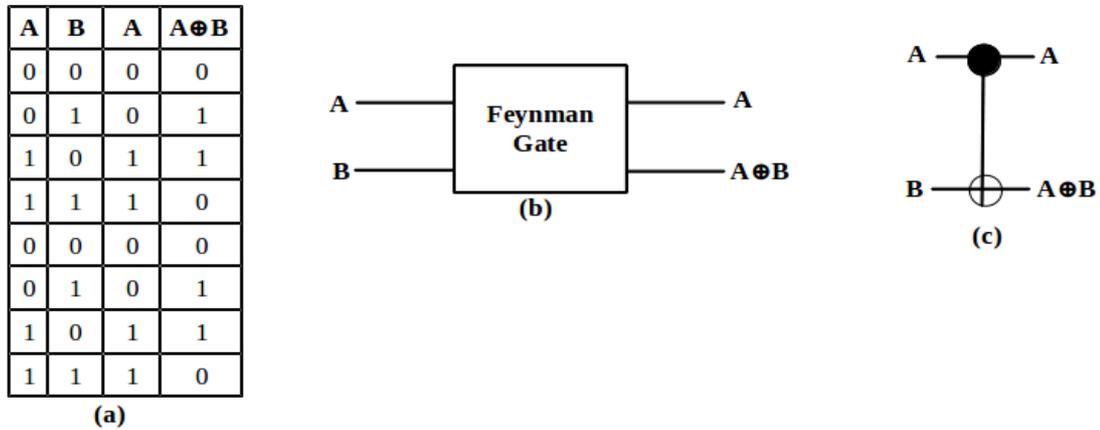

Figure 2.3 Feynman gate (a) Truth Table; (b) Block Diagram; (c) Quantum Implementation

➢ **Toffoli Gate:** A Toffoli Gate (TG) is a 3-input 3-output two-through reversible gate as depict in Figure 2.4(b). Two-through means two of its output signals are the same as input signals with mapping (A, B, C) to (P = A, Q = B, R = A·B $\oplus$ C), where A, B, C are input signals and P, Q, R are output signals, respectively. Toffoli gate is one of the most common reversible gates and has quantum cost of 5. The quantum cost of Toffoli gate is 5 as it requires 2 CNOT gates, 2 Controlled-V gates, 1 Controlled-$V$+ gate realize it. This gate can be used to realize a 2-input reversible AND function by setting C as a constant 0.

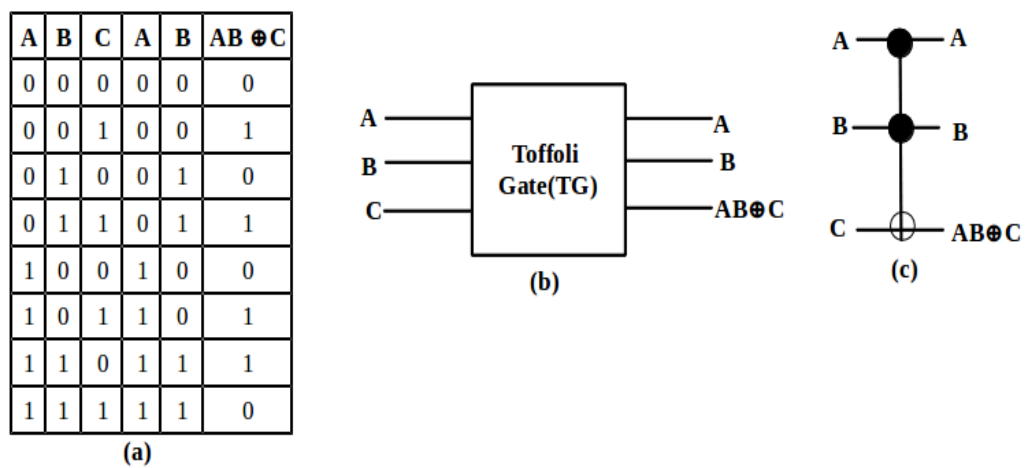

Figure 2.4 Toffoli gate (a) Truth Table; (b) Block Diagram; (c) Quantum Implementation

➢ **Peres Gate:** A Peres gate is a 3×3 one- through reversible gate with mapping (A, B, C) to (P = A, Q = A ⊕ B, R = (A · B) ⊕ C), where A, B, C are the input signals and P, Q, R the output signals, respectively. Figure 2.5(b) shows the block diagram of Peres gate. Quantum cost of Peres gate is 4. In the existing literature, among the 3×3 reversible gate, Peres gate has the minimum quantum cost.

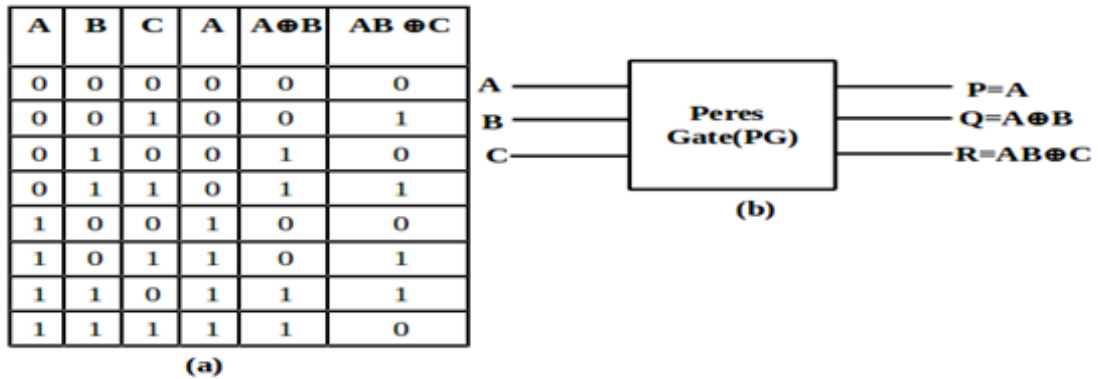

Figure 2.5  Peres gate (a) Truth Table; (b) Block Diagram

### 2.2.3 Reversible Circuit

Reversible circuits are composed of several reversible logic gates. For the realization of reversible logic circuits several synthesis methods have been addressed in the literature [14]. A reversible library is a set of logic gates used for the development of reversible circuits. Reversible logic circuits perform undo operation when run backward. A reversible circuit with two Toffoli gates, one CNOT gate and one NOT gate is as shown in Figure 2.6.

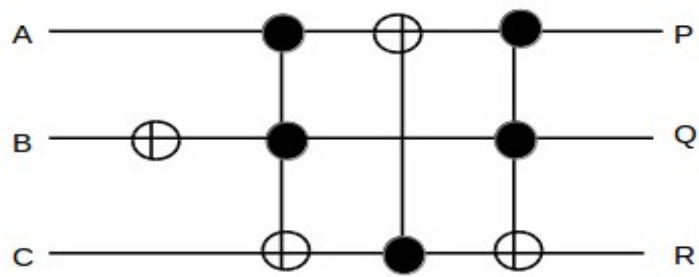

Figure 2.6  A Reversible Circuit

### 2.2.4 Optimization Parameters

The synthesis of sequential reversible logic is carried out by transformation based. After the synthesis process, the optimization of the circuit is required. The optimization process can be started with the synthesis process or after the synthesis. It is also called the post synthesis optimization. The design of any reversible logic circuit should be optimized based upon some necessary parameters. These parameters are explained as follows:

➢ **Gate Counts:** The total number of gates used in a reversible logic circuit is known as Gate Counts.

➢ **Garbage Outputs:** The unwanted or unused outputs which are needed to maintain reversibility of a reversible gate are called as Garbage Outputs. In reversible sequential elements to denote the garbage output we use symbol $g_1$, $g_2$, $g_3$ etc. For example, Feynman gate (FG) is used to perform Exclusive-OR between two inputs. But in that case, one extra output will be generated as well, which is the garbage output.

➢ **Constant Inputs:** Constants are the input lines that are either set to zero or one in the circuit's input side.

➢ **Quantum Cost:** Each reversible logic gate has a cost associated with it called quantum cost [15]. The quantum cost of a reversible gate is the number of $1\times1$ and $2\times2$ reversible gates or quantum logic gates required in its design. The computational complexity of a reversible gate can be represented by its quantum cost. The quantum costs of all reversible $1\times1$ and $2\times2$ gates are taken as unity.

➢ **Hardware Complexity:** The total number of logic operations in a circuit is known as hardware complexity. In hardware complexity the terms are:

$\alpha$ = A two input EX-OR gate calculation

$\beta$ = A two input AND gate calculation

$\delta$ = A NOT calculation

Basically, it refers to the total number of EX-OR, AND & NOT operation in a circuit.

## 2.3  Fault Tolerance in Reversible Logic

Fault-tolerance is the property that enables a system to operate accurately in the presence of the failure of one or more of its components. Fault tolerance in reversible circuit reflects robustness of the system. Fault tolerant systems are capable for detection and correction of faults. If the logic circuit itself is made of fault tolerant components, then the detection and correction of faults in circuit become cheaper, easier and simple. To achieve fault tolerance, the first step is to identify occurrence of fault. To detect the happening of fault, parity preserving technique has been addressed in literature. Any fault that affects only one signal is detectable at the circuit's primary outputs in parity preserving reversible circuits [10]. Hence, using parity preserving reversible logic circuits we can incorporate fault tolerance in reversible computing. Conservative logic is an alternative for achieving fault tolerance in reversible computing and has capability of multi bit fault detection but it requires high cost and complex resultant reversible circuit.

Fault detection in reversible logic circuits can be incorporated using fault tolerant reversible gates. Fault tolerant reversible gates & reversible circuits including parity-preserving and conservative reversible logic gates are addressed one-by-one as follows.

### 2.3.1 Parity-Preserving Reversible Logic Gates

A reversible logic gate is called parity preserving if it preserved parity of input data in their outputs. Means, EX-OR of all input signals is equal to EX-OR of all output signals. In literature following parity-preserving reversible logic gates reported:

➢ **Fredkin Gate:** Fredkin gate is the only basic reversible logic gate which is parity-preserving. It is a 3×3 reversible logic gate, with mapping (A, B, C) to (P = A, Q = A'B + AC, R = AB + A'C), where A, B, C are the input signals and P, Q, R are the output signals, respectively [8]. In simple words the bottom two signals are swapped if the upper signal has input value 1. Otherwise, no change will occur. Figure 2.7 shows a 3*3 Fredkin gate.

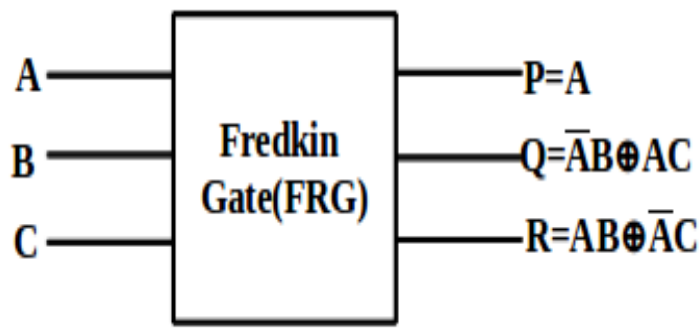

Figure 2.7 Fredkin Gate

Figure 2.8 shows Fredkin gate quantum implementation with quantum cost of 5 [16]. Each dashed rectangles in Figure 2.8 is equivalent to a 2×2 gate and so the quantum cost of each dashed rectangle is unit [15]. This assumption is considered in Hung et al. [16] for calculation of the quantum cost. Hence Fredkin gate cost consists of 2 dashed rectangles, 2 CNOT gates and 1 Controlled-V gate resulting in its quantum cost as 5.

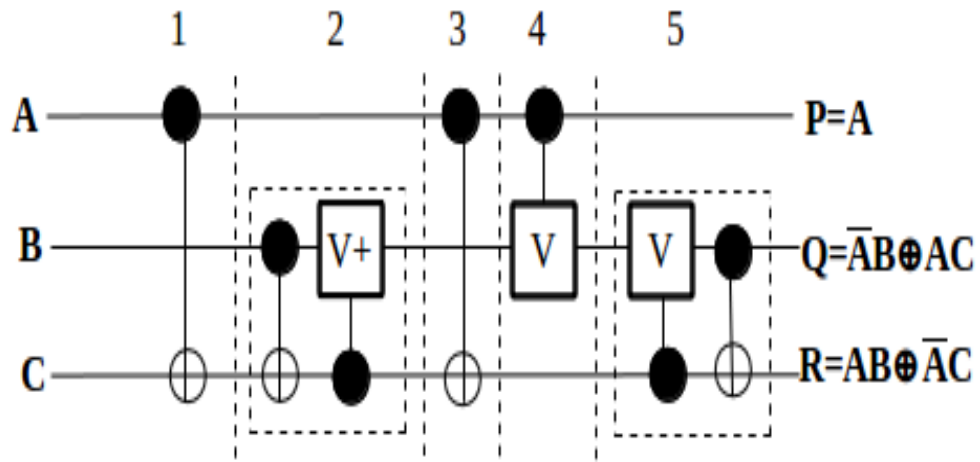

Figure 2.8 Quantum Realization of Fredkin Gate

From truth table shown in table 2.1 it can be verified that Fredkin gate is parity-preserving.

Table 2.1 Truth Table of Fredkin Gate

| A | B | C | P | Q | R |
|---|---|---|---|---|---|
| 0 | 0 | 0 | 0 | 0 | 0 |
| 0 | 0 | 1 | 0 | 0 | 1 |
| 0 | 1 | 0 | 0 | 1 | 0 |
| 0 | 1 | 1 | 0 | 1 | 1 |
| 1 | 0 | 0 | 1 | 0 | 0 |
| 1 | 0 | 1 | 1 | 1 | 0 |
| 1 | 1 | 0 | 1 | 0 | 1 |
| 1 | 1 | 1 | 1 | 1 | 1 |

➢ **Feynman Double Gate:** Feynman Double Gate (F2G) is a 3×3 parity-preserving reversible logic gate [10]. The last two signal lines performed EXOR with value of signal line A. Here A is the control line and B and C are the target lines. It is one through gate with mapping (A, B, C) to (P = A, Q=A ⊕ B, R= A ⊕ C), where A, B, C are the input signals and P, Q, R are the output signals, respectively.

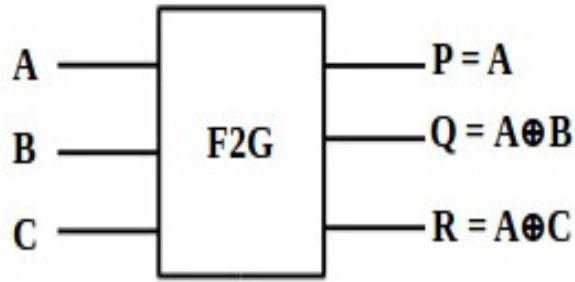

Figure 2.9  Feynman Double Gate

Figure 2.9 shows block diagram of F2G gate. Table 2.2 shows the truth table of parity-preserving F2G gate.

Table 2.2   TRUTH TABLE OF FEYNMAN DOUBLE GATE

| A | B | C | P | Q | R |
|---|---|---|---|---|---|
| 0 | 0 | 0 | 0 | 0 | 0 |
| 0 | 0 | 1 | 0 | 0 | 1 |
| 0 | 1 | 0 | 0 | 1 | 0 |
| 0 | 1 | 1 | 0 | 1 | 1 |
| 1 | 0 | 0 | 1 | 1 | 1 |
| 1 | 0 | 1 | 1 | 1 | 0 |
| 1 | 1 | 0 | 1 | 0 | 1 |
| 1 | 1 | 1 | 1 | 0 | 0 |

### 2.3.2   Conservative Reversible Logic Gates

A reversible logic gate is called conservative if the number of logical ones (Hamming weight) of its input lines equals the number of logical ones of their output [19]. Hence, detection of multiple bit errors is possible at the circuit outputs. In literature only one gate is conservative that is, Fredkin gate. From Table 2.1 it can be proved that the Fredkin gate is conservative.

### 2.3.3  Parity-Preserving Reversible Circuits

A reversible circuit is parity preserving if each gate of the circuit is parity preserved. In this type of circuit input parity of input data is maintained throughout the computation. In communication and digital system fault tolerance is achieved by parity preservation. In sequential circuit design, parity checking is one of the widely used error detection methodology. . Any fault that affects maximum single line is readily detectable at the primary outputs in parity preserving reversible logic circuits [10]. Figure 2.10 shows a fault tolerant (parity-preserving) D latch proposed in [20]. This reversible D flip-flop is composed of one parity-preserving Fredkin gate and one parity-preserving F2G gate.

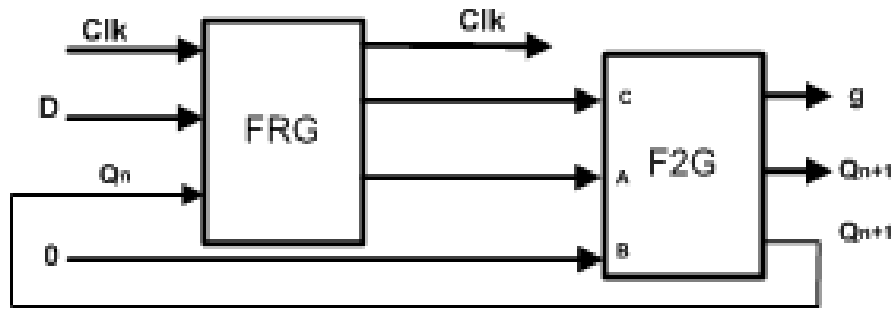

Figure 2.10  Parity Preserving Reversible Sequential Circuit

### 2.3.4  Conservative Reversible Circuits

A reversible logic circuit is called conservative if it preserves the number of logical ones in each input-output pair [19]. These circuits are highly cost constraint and complex in nature. Figure 2.11 shows a fault tolerant (Conservative) T flip-flop proposed in [21]. This reversible T flip-flop is composed three conservative Fredkin gates.

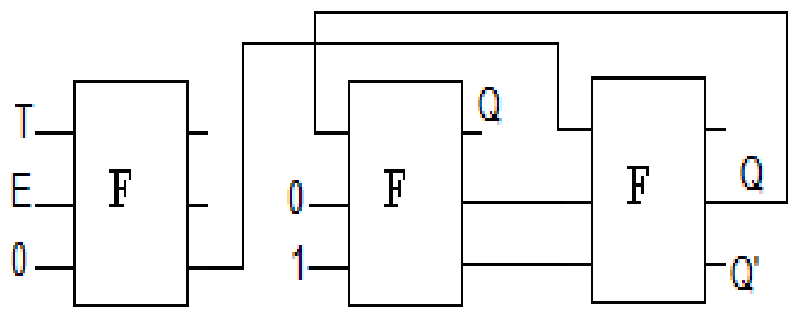

Figure 2.11  Conservative Reversible Sequential Circuit

### 2.3.5  Fault Models in Reversible Logic

Fault is an error occurred in the systems which force system to deviate from its normal behavior. In both traditional and reversible logic circuits the complexity of generating tests for all possible faults in a logic circuit can be minimized through the use of fault models which cover particular set of fault possibilities. The fault models vary according to the type of description that is being considered, which in turn varies according to the level of abstraction [22]. A number of fault models for reversible logic circuits have been addressed in literature for efficient fault diagnosis. Normally fault models can be categorized as single fault models and multiple fault models depending on the number of fault can occur in the system. The description of fault models, reported in literature are detailed as follows:

➢ **Stuck-at Fault Model:** The stuck-at fault [23] of gate causes one of its inputs or outputs to be stuck at either 1 (stuck-at-1) or 0 (stuck-at-0) without considering the correct value of input line or output line. This is a widely used common fault model for traditional logic. According to [22] the stuck-at fault model can be used for reversible computing. For example we have a reversible circuit shown in Figure 2.12. A stuck-at-0 on first output of the first gate gives value 0 to output P without going for considering the input value.

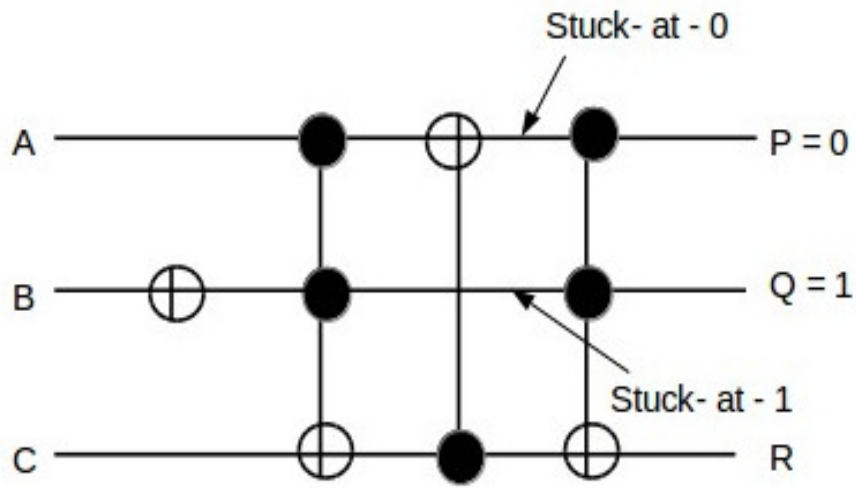

Figure 2.12  Stuck-at- Fault Model

➢ **Bit Fault Model:**  The bit fault model for a reversible logic assumes that one or more lines will have their value altered from the correct value to some incorrect value. Single-bit fault for a reversible logic flips one of the values of its output from 1 to 0 or vice versa. Technological reasons for this alter are not specified, as one assumes that any number of reasons could be the cause for such an alteration. Unlike stuck-at fault model, this model is depending on the input value. This fault model is reported in [22]. Figure 2.13 shows the circuit with a bit fault.

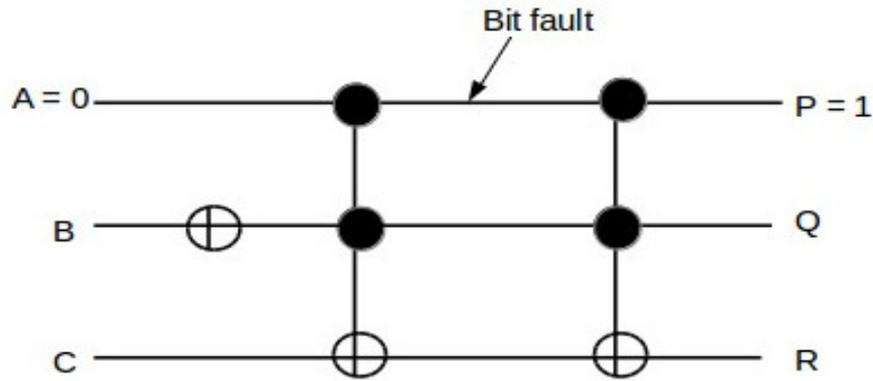

Figure 2.13 Bit Fault Model

### 2.3.6 Fault Testing Approaches

Fault testing in reversible logic circuit is commonly used for detection of faults occurred in the circuit. According to [24] fault testing can be performed online and offline. Online approaches for fault detection represent testing of design in their normal operation while offline requires extra overhead to detect fault. Offline testing approaches use test vectors to detect faults in the circuit whereas online testing approaches detect fault in normal circuit operation. Here we discussed each of fault testing approaches in brief one by one.

➤ **Offline Testing:** In Offline testing approach the circuit will be taken out of normal operations and can be tested by applying a number of test vectors to the circuit for which the correct output values for the circuit are known. Thus a key element in offline testing approaches for a given fault model is the computation of test sets that are complete for the model under consideration. Input vector that is used for testing a circuit offline for fault detection known as test vector. A set of test vectors is known as test set. A test-set is complete if it is capable in detection of all faults in the fault set S, and such a test set is minimal if it contains the most fewest possible vectors [22]. Sometimes additional modification in circuitry required, in which case the approach referred to as a design-for-test (DFT) approach.

➤ **Online Testing:** In online testing approaches fault can be identified while the circuit is operating normally. It is performed during normal operation of the circuit. This may require the addition of circuitry to enable the detection of faults while the circuit is being used in normal operation. Thus offline testing approach requires extra overhead to detect fault in circuit.

## 2.4  Related Work

A comprehensive and in depth evaluation of previously reported work in literature is presented in this section. Researchers have addressed the design of reversible sequential building blocks & circuits. The following subsections describe the existing work in the design of reversible sequential reversible building blocks, reversible circuits and fault tolerant design of reversible sequential building blocks.

### 2.4.1  Reversible Sequential Building Blocks

In existing literature, researchers have addressed the design of reversible sequential building blocks. Reversible flip-flops are the most significant and basic memory elements that will be the target building block of memory for the realization of reversible sequential circuits. In 2005, the first attempt on the design of reversible building blocks was H. Thapliyal  et al. [25]. In this work, the Fredkin, Feynman and New Gate was used as AND, NOT and NOR Gate respectively. In the designing of reversible flip-flop, the conventional design of a flip-flop was used. The proposed design of D flip-flop has 7 reversible gates, 8 garbage outputs. The design of R-S flip-flop in this work has 6 reversible gates, 8 garbage outputs. The proposed design of J-K flip-flop has 10 reversible gates, 12 garbage outputs. The proposed design of T flip-flop has 10 reversible gates, 12 garbage outputs. This work was first attempt and required further investigation in order to optimize these proposed designs. Only two optimization parameters were considered by researcher in this work to analyse their realization of reversible sequential building blocks.

In 2006, J. E. Rice [26] has proposed a new reversible implementation for a reversible R-S latch. All reversible flip-flops (except R-S) have been realized using R-S latch in this work. For the designing of reversible flip-flops, Toffoli and Feynman Gate were used as CCNOT and CNOT gate respectively. S. K. S. Hari et al. [27] have addressed reversible flip-flops by using basic reversible Fredkin and Feynman gates in 2006. Proposed realization of reversible sequential building blocks were analyzed in terms of gate count and garbage output. The reversible flip-flops were proposed by A. Banerjee et al. [9] in 2007. For the construction of reversible flip-flops, Toffoli gate, Feynman and NOT were used.

In 2008, a novel concept on the designing of reversible flip-flops was proposed by Min-Lun Chuang et al. [28]. First, reversible latches are proposed and with help of these latches flip-flops were realized in this work. This work was an improvement over the proposed designs of reversible building blocks in term of optimization parameter by H. Thapliyal. The proposed realizations were optimized in terms of gate counts, garbage outputs which were 5 to 7 and 2 to 4 respectively. N.M. Nayeem et al. [29] have reported the D flip-flop by using Fredkin and Feynman gate in 2009. The optimized realization was compared with previous work on the basis of gate counts, garbage outputs and quantum cost.

H. Thapliyal et al. [30] have addressed the optimized design of sequential reversible circuits such as reversible D flip-flop, T flip-flop, JK flip-flop, R-S flip-flop and reversible master-slave flip-flops in 2010. In this work, designs of sequential reversible building blocks are optimized for gate count, delay and garbage outputs. V. Rajmohan et al. [31] have reported the realization of reversible D flip-flop in 2011. The realization of D flip-flop has single Sayem gate. Hence, in this work gate count was one to realize D flip-flop.

### 2.4.2 Reversible Shift Registers & Shift Counter

In current literature, researchers have proposed the designs of sequential reversible shift registers. In 2006, H. Thapliyal et al. [32] have reported complex sequential reversible circuit. In this work reversible SIPO shift register is realized from reversible D flip-flop and Feynman gate with common clock input.

N.M. Nayeem et al. [29] have reported efficient and optimized designs of shift registers in 2009. The optimized designs were compared with previous work on the basis of gate counts, garbage outputs and quantum cost. In 2011, the reversible D flip-flop was addressed by V. Rajmohan et al. [31]. Proposed D flip-flop was used in proposing the design of serial and parallel shift register. The proposed design of shift registers were highly optimized as compared with previous work.

### 2.4.3 Fault Tolerance design of Sequential Building Blocks

In literature a few of the early reported works are related to fault tolerant design of reversible building blocks and sequential circuits. In 2006, the first attempt of fault detection in reversible circuits through parity preservation characteristic was addressed by Behrooz Parhami [10]. In 2010, the fault tolerant D flip-flop and shift registers were addressed by Majid Haghparast et al. [33]. In this work, proposed fault tolerant (parity preserving) D flip-flop requires three F2G and two Fredkin gate. Fault tolerant D flip-flop used to realize fault tolerant shift register.

Ali Hatam et al. [20] have extended work of Haghparast et al. [33] and reported n-bit fault tolerant shift register realization in 2011. H. Thapliyal et al. [34] have addressed the design of testable sequential reversible circuits such as reversible positive enable D flip-flop, a negative enable D flip-flop and reversible master-slave D flip-flop in 2012. In this work, realizations of sequential reversible building blocks are testable for stuck-at-faults. In 2012, fault tolerant reversible J-K and D flip-flop were proposed by Lafifa Jamal et al. [35] using basic reversible Fredkin and Double Feynman gates. Proposed realization of reversible flip-flop were analyzed in terms of gate count, garbage output and quantum cost. In 2013, J. E. Rice [22] has discussed overview of fault models and online and offline testable approaches to detect the fault in reversible circuits.

### 2.4.4  Survey Extraction

From a careful survey of the existing works on reversible sequential circuits, it can be summarized that most of these work considered the optimization of number of reversible gates, constant input and garbage outputs, while ignoring the important parameters of quantum cost and hardware complexity. We have observed that the realizations of reversible sequential building blocks and reversible sequential circuits can be further optimized in terms of all important optimization parameters (gate count, constant input, garbage output, and quantum cost and hardware complexity). Apart from this, it is also observed that the optimization parameters like quantum cost and garbage outputs were very high in realization of fault tolerance sequential building blocks. We observe that these optimization parameters can further be reduced.

In existing literature fault tolerance designs for reversible sequential circuit are based solely on parity preservation. To the best of our knowledge, the online testing of faults in reversible sequential circuits is not addressed in the literature. Hence, online testable design of reversible sequential building blocks can be proposed to detect faults online. The next subsequent chapters of this thesis report will explain our proposed work in detail.

DESIGN OF REVERSIBLE GATE & SEQUENTIAL BUILDING BLOCKS

This chapter describes the proposed optimized realization of sequential reversible building blocks and circuits. Reversible flip-flops are the most significant and basic memory elements that will be the target building block of memory for the forthcoming computing devices. The proposed design has improved optimization parameters. In this chapter, a new parity preserving reversible gate is proposed and this gate with basic reversible gate is used to realize the reversible sequential building blocks such as positive level D flip-flop, R-S flip-flop, JK flip-flop and T flip-flop. We have also proposed a quantum realization of our proposed parity preserving reversible gate. The low cost realization of reversible Serial in and Parallel Output (SIPO) shift register, Parallel in and serial Output (PISO) shift register and shift register counter (Johnson counter) is also proposed by proposed realization of flip-flop with basic reversible gates. To cover these proposed designs, sections are includes in this chapter are as follows:

➢ Proposed Reversible Parity Preserving Gate
➢ Quantum Realization of Proposed Gate
➢ Proposed Reversible Positive Level Flip-Flops
➢ Design of SIPO & PISO Shift Registers
➢ Design of Shift Register Counter (Johnson Counter)

## 3.1 Reversible Parity Preserving Gate

We propose a new 4×4 parity preserve reversible circuit, called Pareek gate. The block diagram of the proposed gate is shown in Figure 3.1.

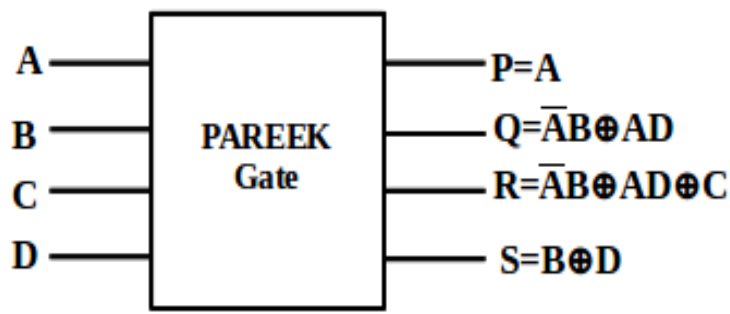

Figure 3.1   Proposed Parity Preserving Reversible Pareek Gate

The truth table of proposed parity preserving Pareek Gate is shown in Table 3.1. The output P (=A) is copied directly from input A, this input to output line is called control line where as other lines are called target lines. The gate produces three outputs, namely, Q, R and S on target lines as defined in Figure 3.1. The outputs are verified manually through the truth table.

Table 3.1    TRUTH TABLE OF THE PROPOSED REVERSIBLE PAREEK GATE

| Input | | | | Output | | | |
|---|---|---|---|---|---|---|---|
| A | B | C | D | P | Q | R | S |
| 0 | 0 | 0 | 0 | 0 | 0 | 0 | 0 |
| 0 | 0 | 0 | 1 | 0 | 0 | 0 | 1 |
| 0 | 0 | 1 | 0 | 0 | 0 | 1 | 0 |
| 0 | 0 | 1 | 1 | 0 | 0 | 1 | 1 |
| 0 | 1 | 0 | 0 | 0 | 1 | 1 | 1 |
| 0 | 1 | 0 | 1 | 0 | 1 | 1 | 0 |
| 0 | 1 | 1 | 0 | 0 | 1 | 0 | 1 |
| 0 | 1 | 1 | 1 | 0 | 1 | 0 | 0 |
| 1 | 0 | 0 | 0 | 1 | 0 | 0 | 0 |
| 1 | 0 | 0 | 1 | 1 | 1 | 1 | 1 |
| 1 | 0 | 1 | 0 | 1 | 0 | 1 | 0 |
| 1 | 0 | 1 | 1 | 1 | 1 | 0 | 1 |
| 1 | 1 | 0 | 0 | 1 | 0 | 0 | 1 |
| 1 | 1 | 0 | 1 | 1 | 1 | 1 | 0 |
| 1 | 1 | 1 | 0 | 1 | 0 | 1 | 1 |
| 1 | 1 | 1 | 1 | 1 | 1 | 0 | 0 |

It is observed that the parity of the input bits is equal to the parity of the output bits in each row of the Table 3.1. Hence, the gate also preserves the parity. This characteristic can further be used in fault tolerant design of reversible sequential circuit. However, we propose the low cost design of reversible sequential building blocks using this gate.

## 3.2 Quantum Realization of Proposed Gate

The quantum cost of a reversible gate is the number of 1×1 and 2×2 reversible gates or quantum logic gates required in its design. The computational complexity of a reversible gate can be represented by its quantum cost. The quantum costs of all reversible 1×1 and 2×2 gates are taken as unity. Any reversible gate can be realized using the 1×1 NOT gate, and 2×2 reversible gates such as Controlled-V and Controlled-V$^+$ and the Feynman gate which is also known as the Controlled NOT gate (CNOT). Thus, it can said that the quantum cost of a reversible gate can be calculated by counting the numbers of NOT, Controlled-V, Controlled-V$^+$ and CNOT gates required in its implementation. The quantum cost of proposed reversible gate is calculated by an optimization algorithm [37].

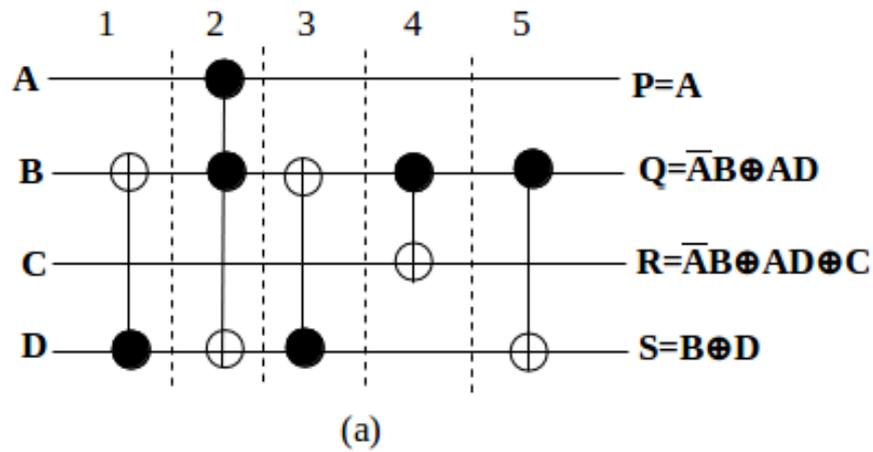

Figure 3.2 (a) Realization of Proposed Gate Using Toffoli and CNOT Gate

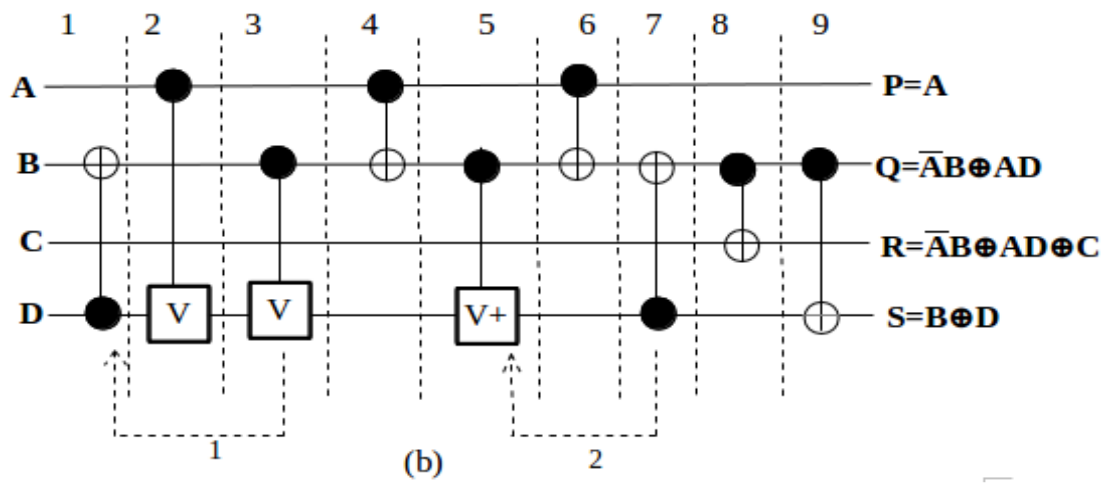

Figure 3.2 (b) Quantum Realization of Proposed Pareek Gate

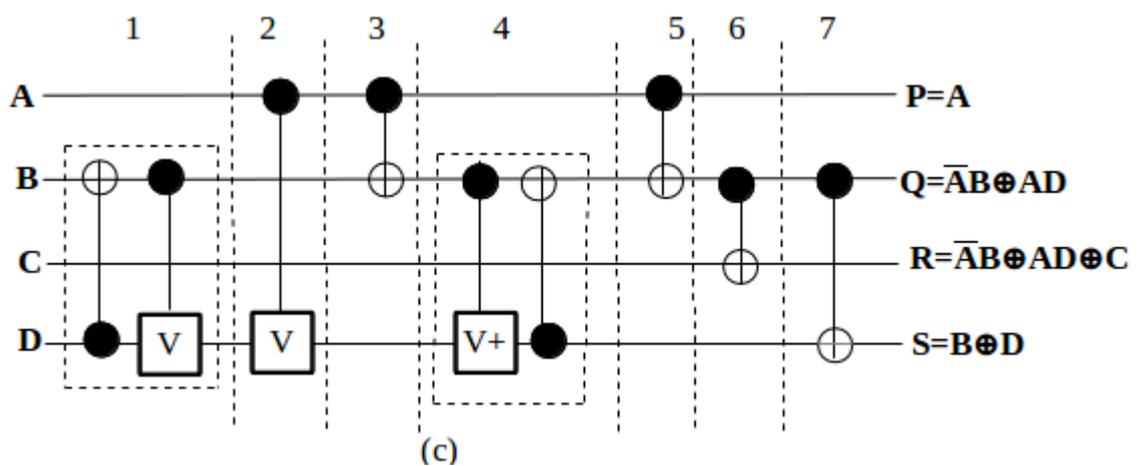

Figure 3.2 (c) Optimized Quantum Realization of Proposed Gate

In Figure 3.2 (a), the proposed Pareek gate is realized using one Toffoli gate and four CNOT gates. Then, Toffoli and CNOT gates are substituted by quantum primitives and moving rule is applied (the movements are shown by arrows), so its direct linear cost is (1×5) + (4×1) =9, which is shown in Figure 3.2 (b). New gates are introduced (dashed boxes) in Figure 3.2 (c) to yield quantum cost of Pareek gate as 7.

## 3.3 Reversible Level Triggered Flip-Flops

A flip-flop is a circuit that has two stable states and can be used to store state information. It is the basic storage element in sequence logic. In positive level triggered flip-flop whenever the clock input is high (logic 1) the data present in the data line is stored in the flip-flop. In negative level triggered flip-flop whenever the clock input is low (logic 0) the data present in the data line is stored in the flip-flop. In this section, first we discuss the design methodology for synthesis of reversible sequential building blocks. The design of reversible positive level flip-flops is proposed in this section. Reversible negative level flip-flop can be design in similar way.

### 3.3.1   Design Methodology for Synthesis of Sequential Reversible Building Blocks

The output expressions of the reversible logic gates are used as the templates for mapping the characteristic expression of the flip-flop into an equivalent reversible realization. For example, in the characteristic expression of a flip-flop, suppose we have an expression as A'B + AC, it can be easily mapped with the template of the output expression of the Fredkin gate, and hence Fredkin gate can be used to synthesis this characteristic expression of supposed flip-flop. The proposed design procedure is illustrated here with the design of the reversible positive level triggered D flip-flop as an example circuit.

**[1] Step 1:** Generate templates of the output expression of the basic reversible gates and sequential reversible gate as follows:

- a.   Fredkin gate: FRG(A, B, C) = A'B + AC or  FRG(A, B, C) = AB + A'C
- b.   Peres gate: PG(A, B, C) = A · B $\oplus$ C
- c.   Toffoli gate: TG(A, B, C) = A · B $\oplus$ C
- d.   Feynman gate: FG(A, B) = A $\oplus$ B
- e.   NOT gate: NOT(A) = A'
- f.   Feynman Double gate: F2G(A, B, C) = A $\oplus$ B or F2G(A, B, C) = A $\oplus$ C

g. Pareek gate: Pareek(A, B, C, D) = A'B $\oplus$ AD or Pareek(A, B, C, D) = A'B $\oplus$ AD $\oplus$ C

**[2] Step 2:** Derive the characteristic expression of the flip-flop.

The characteristic equation of the R-S flip-flop can be derived as $Q_{t+1}$ = D.CLK + CLK'.$Q_t$ Where D is the inputs to D flip-flop, $Q_t$ is the previous output, CLK is the clock signal and $Q_{t+1}$ is the current output.

**[3] Step 3:** From the templates in Step 1, find the template which exactly maps the characteristic equation with minimum quantum cost and derive the minimum number of garbage outputs needed to convert the characteristic expression as a reversible function. For D flip-flop Pareek gate is exactly maps characteristic equation with two garbage outputs.

**[4] Step 4:** Keep off fan-out in the resultant reversible logic circuit by efficiently using Feynman gates. Further maintain low cost in terms of garbage outputs by carefully utilizing the output expressions of intermediate stages.

### 3.3.2 Reversible Level Triggered D Flip - Flop

The reversible D flip-flop has the characteristic equation $Q_{t+1}$ = D.CLK + CLK'.$Q_t$. In the proposed realization CLK is the clock. When the clock is 1 the input D is passed to the output that is $Q_{t+1} = D$, when the clock is zero the output remains in the previous state. To realize the reversible D flip-flop with Q output, we are using one Pareek Gate. The proposed D flip-flop has one Gate, one constant input, two garbage outputs and 7 quantum cost.

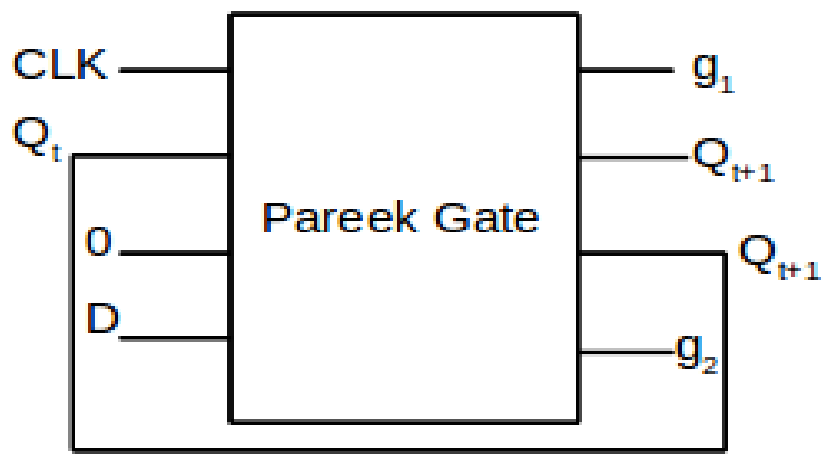

Figure 3.3  Proposed Reversible Positive Level D Flip-Flop

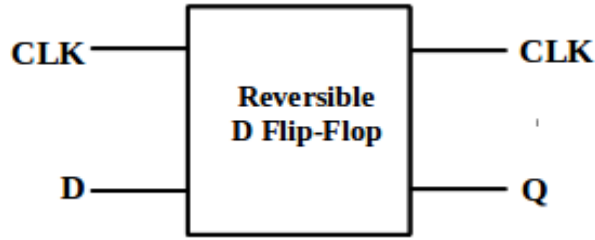

Figure 3.4 Block Diagram of Proposed Reversible Positive Level D Flip-Flop

The realization of reversible positive level D flip-flop with Q output is shown in Figure 3.3 and the corresponding block diagram is shown in Figure 3.4. Due to the proposed parity preserving gate, the realization of reversible D flip-flop is also parity preserving. Hence, proposed design of reversible positive level D flip-flop is fault tolerance.

### 3.3.3 Reversible Level Triggered R-S Flip - Flop

The characteristic equation of the R-S flip-flop can be written as $Q_{t+1} = (S.CLK + (R.CLK)'Q_t)$. Figure 3.5 shows the proposed design of reversible R-S flip-flop, in which the NOT gate and Toffoli gate is used to produce $S + R'Q_t$. The output $S + R'Q_t$ is passed to the Pareek gate and finally to the Feynman Gate to realize the SR flip-flop.

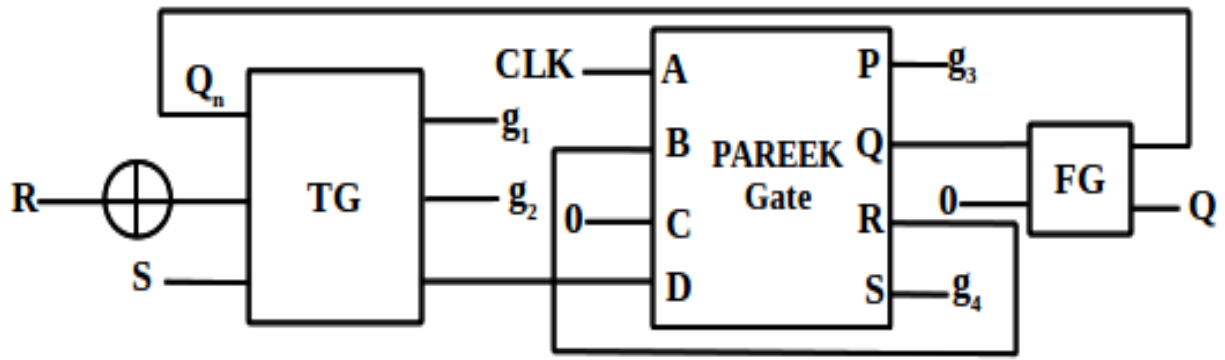

Figure 3.5 Proposed Reversible Positive Level R-S Flip-Flop

The proposed design has one Pareek gate, one NOT gate, one Toffoli and one Feynman gate. It has 2 constant inputs, 4 garbage outputs, 4 gate counts and 13 quantum cost.

### 3.3.4 Reversible Level Triggered JK Flip - Flop

A J-K flip-flop is a refinement of the R-S flip-flop in that the indeterminate state of the R-S type is defined in the J-K type. The JK flip-flop switches to its complement state, whenever input values are applied to both J and K simultaneously. The characteristic equation of the JK flip-flop can be written as $Q_{t+1} = (JQ_t' + K'Q_t).CLK + CLK'Q_t$. Figure 3.6 shows the proposed design of reversible JK flip-flop, in which the NOT gate and Fredkin gate is used to produce $JQ_t' + K'Q_t$. The resultant output $JQ_t' + K'Q_t$ is passed to the Pareek Gate and finally to the Feynman gate to avoid fan-out.

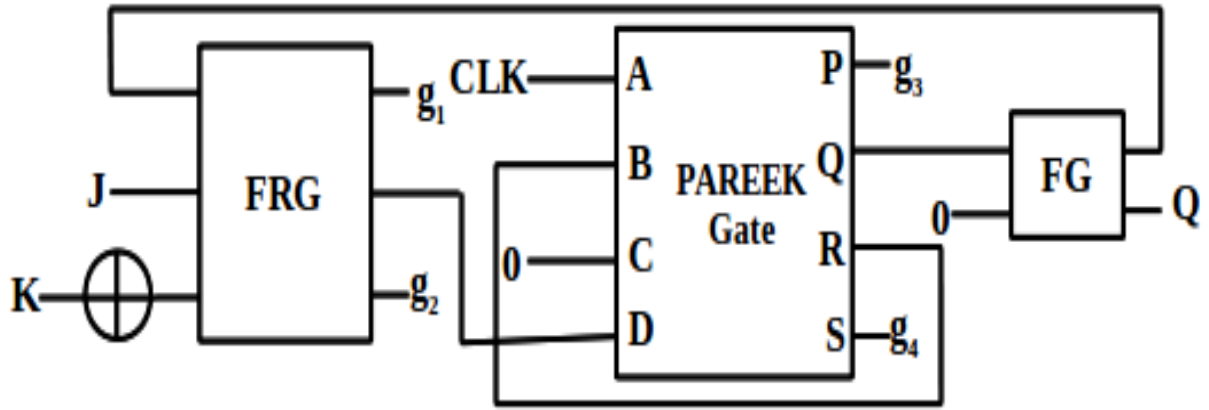

Figure 3.6   Proposed Reversible Positive Level JK Flip-Flop

The proposed design has one Pareek Gate, one Fredkin, one NOT and one Feynman gate. It has 2 constant inputs, 4 garbage outputs, 4 gate counts and 13 quantum cost.

### 3.3.5   Reversible Level Triggered T Flip - Flop

The T flip-flop is a single-input version of the J-K flip-flop. The characteristic equation of the T flip-flop can be written as $Q_{t+1} = (T \oplus Q_t).CLK + CLK'Q_t$. Figure 3.7 shows the proposed design of reversible T flip-flop. The design has one Pareek gate and one Feynman gate.

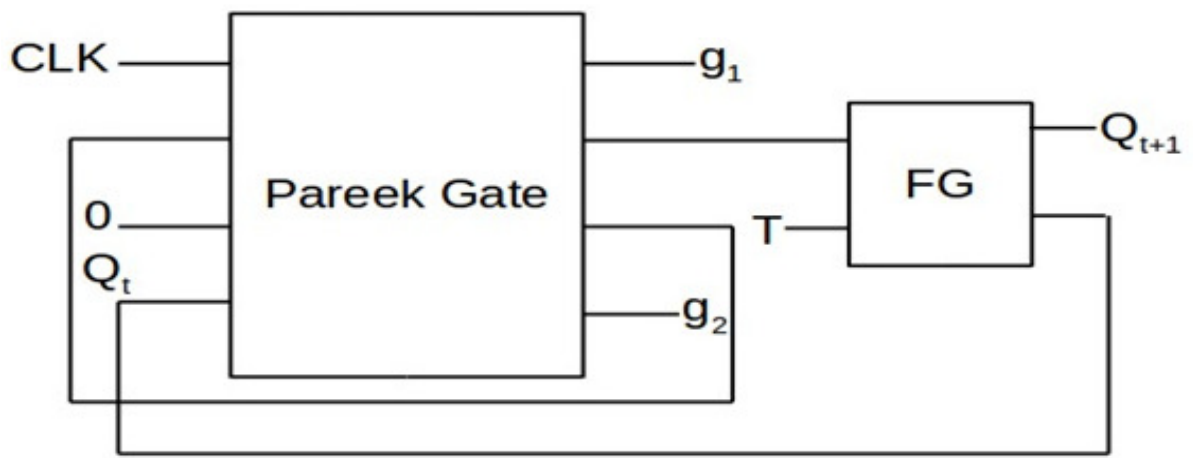

Figure 3.7   Proposed Reversible Positive Level T Flip-Flop

The proposed reversible design of positive level triggered T flip-flop has 1 constant inputs, 2 garbage outputs, 2 gate counts and 8 quantum cost.

## 3.4 Design of Reversible Shift Registers & Shift Register Counter

The shift register is an indispensable functional device in a digital system. A register capable of shifting binary information either to the right or to the left is called shift register. In a shift register, the flip-flops are connected in such a way that the bits of a binary number are entered into the shift register, shifted from one position to another and finally shifted out. Shift registers are used in digital systems for temporary storage of information, data manipulation, transferring and counting circuits. Shift register can be arranged to form counters. Shift register counters use feedback, whereby the output of the last flip-flop in the shift register is connected back to the first flip-flop.

This section provides the reversible design of Serial In and Parallel Output (SIPO) shift register, Parallel In and Serial Output (PISO) shift register and shift register counter by using our proposed reversible gate.

### 3.4.1 Reversible SIPO Shift Register

A 4-bit Serial in parallel out shift register consists of one serial input, and outputs are taken from all the flip-flops parallel. In this register, data is shifted in serially but shifted out in parallel. The reversible design of SIPO shift register using proposed D flip-flop is shown in Figure 3.8.

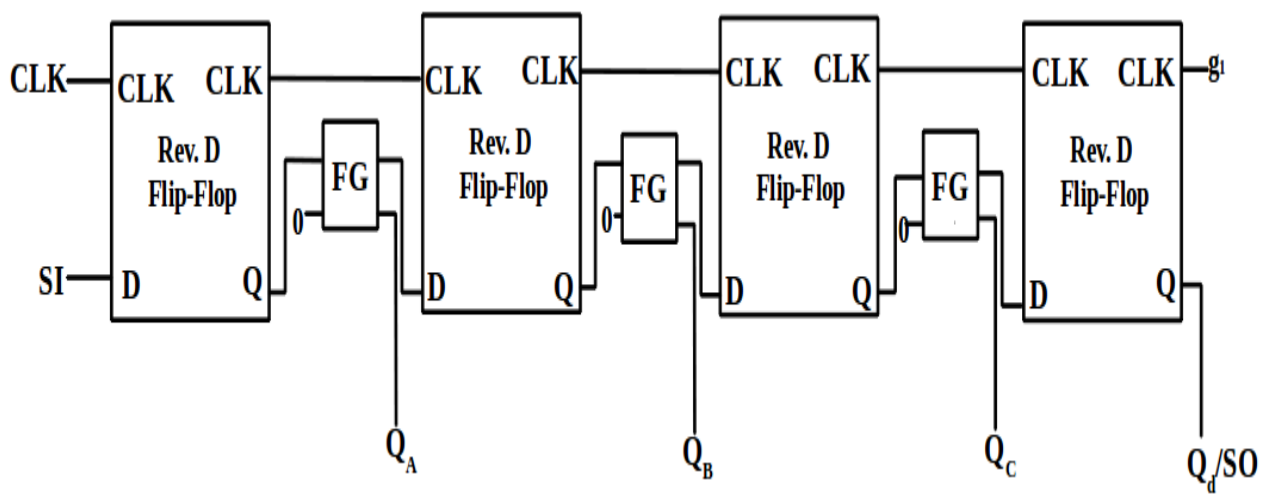

Figure 3.8 Proposed Reversible SIPO Shift Register

The serial input is provided to the SI input of the reversible left-most flip-flop while the outputs $Q_A$, $Q_B$, $Q_C$, $Q_D$ are available in parallel from the Q output of the flip- flops.

### 3.4.2 Reversible PISO Shift Register

In PISO, the bits are entered simultaneously into their respective flip-flops rather than a bit-by-bit basis on one line. The reversible implementation of PISO shift register using proposed reversible D flip-flop is shown in Figure 3.9. According to the conventional design of shift register, the functions are controlled by the clock signal CLK.

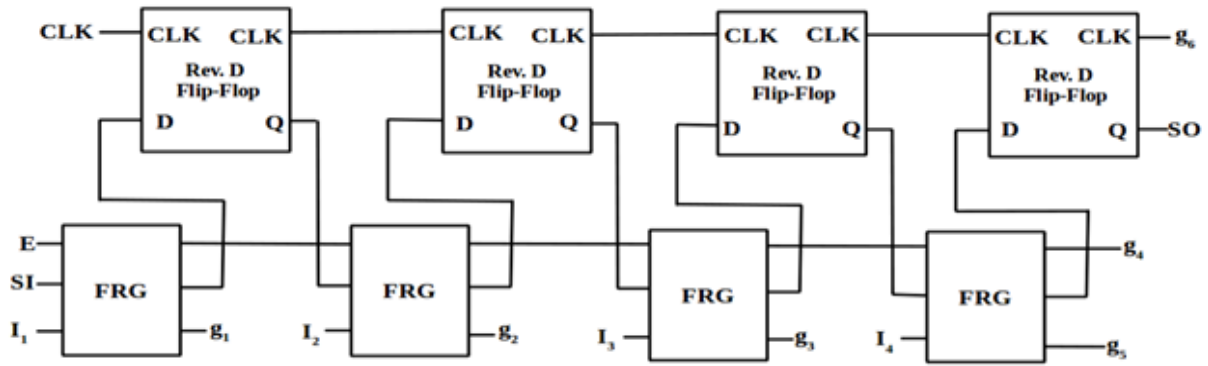

Figure 3.9   Proposed Reversible PISO Shift Register

When clock signal CLK is high, the inputs $I_1$, $I_2$, $I_3$ and $I_4$ are loaded in parallel into the register coincident with next clock pulse. When CLK is low, Q output of D flip-flop is shifted to the right by means of Fredkin Gate. It allows bits are entered simultaneously into their respective flip-flops.

### 3.4.3 Reversible Shift Register Counter (Johnson Counter)

In shift counter, the inverted true output (Q) of the last flip-flop is connected back to the serial input of the first flip-flop. Figure 3.10 shows reversible design of shift counter using proposed reversible D flip-flop.

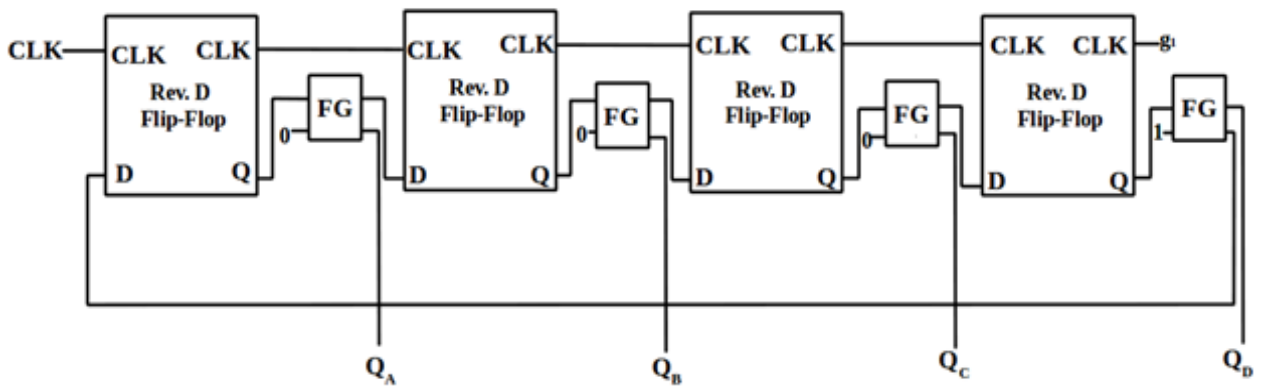

Figure 3.10  Proposed Reversible Shift Register Counter

The output of each reversible flip-flop (Q) is connected to the D input of the next stage. However, the inverted output of the last flip-flop, i.e. $Q_D$ is connected to the D input of the first flip-flop.

The above realizations of reversible sequential blocks and sequential circuits clearly validate our motivation and show that proposed design of sequential building blocks are very cost effective compare to existing design in terms of optimization parameters.

# FAULT TOLERANT & TESTABLE DESIGN OF SEQUENTIAL BLOCKS

This chapter illustrates fault tolerant & testable realization of sequential reversible building blocks. Parhami [2006] proved that, if a circuit has same parity of primary inputs and primary outputs then single bit fault detection is possible. Reversible Circuit constructs with parity preserving gates maintain parity of primary inputs and primary outputs; hence it has single bit fault detection capability. This chapter describes realization of fault tolerance design of reversible sequential building blocks by proposed parity preserving Pareek gate and existing parity preserving gates. Further in this chapter, we elaborate offline testable design of D flip-flop for detect any single stuck at fault and online testable design of D flip-flop to detect bit fault at output. The next sections of this chapter are described as follows:

➢ Fault Tolerant Design of Sequential Reversible Building Blocks
➢ Offline Testable Design of D Flip-Flop
➢ Online Testable Design of D Flip-Flop

## 4.1 Fault Tolerant Design of Sequential Reversible Building Blocks

In this section, we are proposing the realization of fault tolerant sequential reversible building blocks such as positive level triggered D flip-flop, negative level triggered D flip-flop, master-slave flip-flop, Double Edge Triggered (DET) D flip-flop, R-S flip-flop, JK flip-flop and T flip-flop.

### 4.1.1 Fault Tolerant (FT) Reversible Positive Level D Flip-Flop

In the design of fault tolerant reversible positive level triggered D flip-flop, two design approaches can be arises:

➢ ***Design of Reversible Positive Level D Flip-Flop With Q Output:***

The fault tolerant reversible positive level D flip-flop has the characteristic equation $Q_{t+1} = D.CLK + CLK'.Q_t$. In the proposed realization CLK is the clock. When the clock is 1 the input D is passed to the output that is $Q_{t+1} = D$, when the clock is zero the output remains in the previous state. To realize the fault tolerant reversible D flip-flop with Q output, we are using one Pareek Gate. The proposed D flip-flop has one Gate, one constant input, two garbage outputs and 7 quantum cost. The realization of fault tolerant reversible positive level Triggered D flip-flop with Q is shown in Figure 4.1.

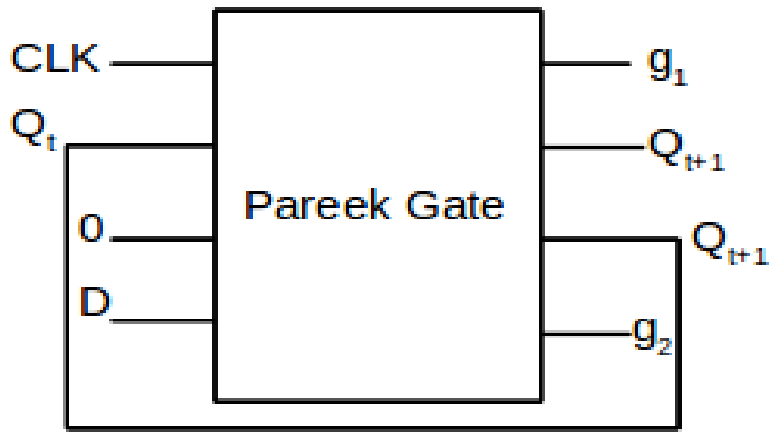

Figure 4.1   Proposed FT Positive level D Flip-Flop With Q Output

➢ *Design of Reversible Positive Level D Flip-Flop With Q and Q' Output:*

The design has shown in Figure 4.1 does not produce the complement output Q', which is required in complex sequential circuits. Very few researchers have shown the outputs Q and Q′ in their work. In this work, we propose a novel design of the fault tolerant reversible positive level D flip-flop that has both the outputs Q and Q'. The design has one Pareek Gate and one F2G as shown in Figure 4.2.

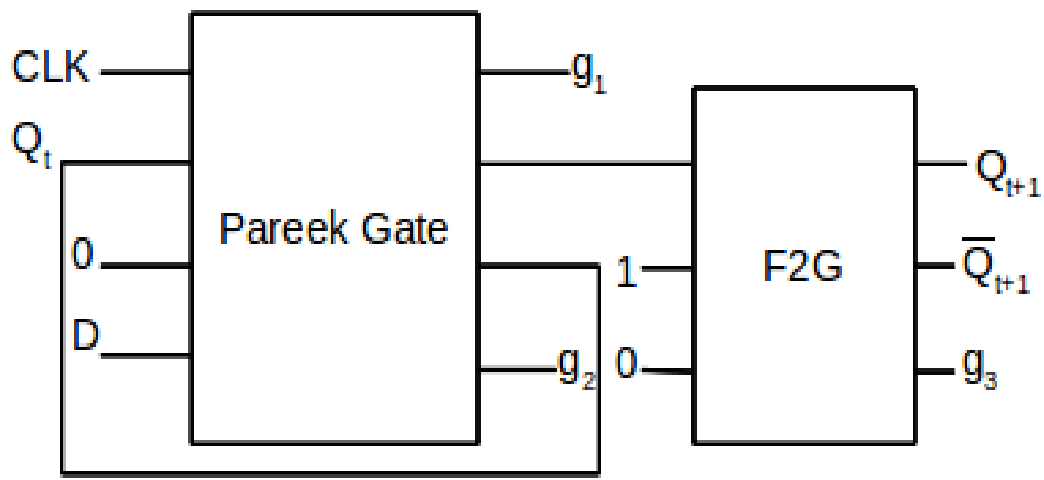

Figure 4.2   Proposed FT Positive level D Flip-Flop With Q and Q' Output

## 4.1.2 Fault Tolerant (FT) Reversible Negative Level D Flip-Flop

In the design of fault tolerant reversible negative level D flip-flop, two design approaches can be arises:

> *Design of Reversible Negative Level D Flip-Flop With Q Output:*

The fault tolerant reversible negative level D flip-flop has the characteristic equation $Q_{t+1} = D.CLK' + CLK.Q_t$. The proposed design passes the input to the output only when the clock is 0 otherwise the flip-flop maintains the same state. To realize the fault tolerant reversible negative level Triggered D flip-flop, we are using one Pareek Gate. The proposed D flip-flop has one Gate, one constant input, two garbage outputs and 7 quantum cost. The realization of fault tolerant reversible negative D flip-flop with Q is shown in Figure 4.3.

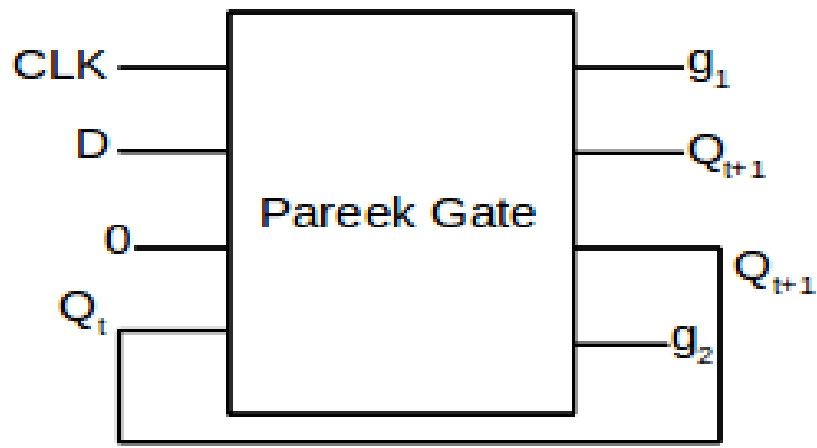

Figure 4.3    Proposed Negative level D Flip-Flop With Q Output

> *Design of Reversible Negative Level D Flip-Flop With Q and Q' Output:*

The design has shown in Figure 4.3 does not produce the complement output Q', which is often required in sequential circuits. Very few researchers have addressed the outputs Q and its complement Q′ in their work.

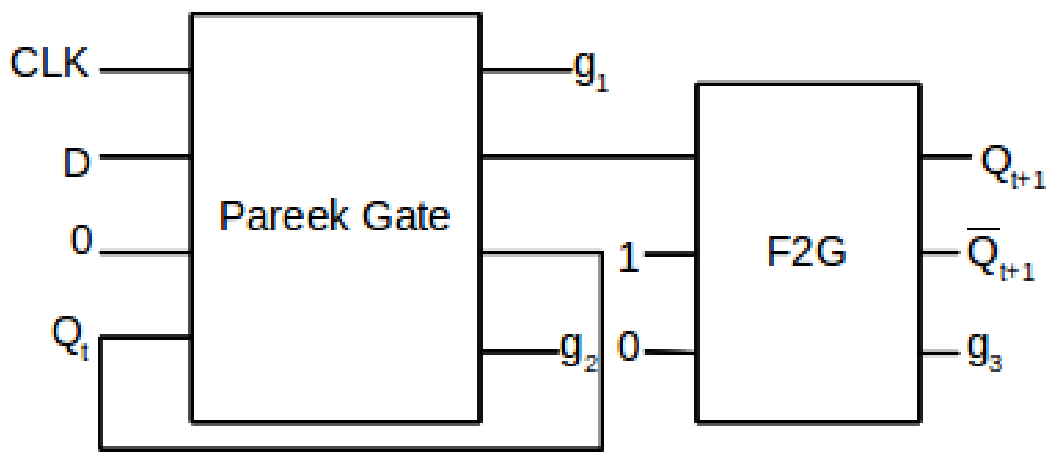

Figure 4.4    Proposed Negative level D Flip-Flop With Q and Q' Output

In this work, a novel design of the fault tolerant reversible negative level D flip-flop that has both the outputs Q and Q' is proposed. The design has one Pareek Gate and one F2G as shown in Figure 4.4.

### 4.1.3 Fault Tolerant Reversible Master-Slave D Flip-Flop

The design of fault tolerant reversible master-slave D flip-flop has positive enabled D flip-flop as the master and the negative enable D flip-flop as the slave. In the master slave flip flop, when clock is high the master passes the input while the slave remains in the previous state and when the clock is low the master flip-flop remains in the storage and the slave passes the output of the master to the output. Thus the master salve flip-flop does not sample at both the positive and negative edge of the clock cycle. The proposed design of fault tolerant reversible master-slave D flip-flop has 2 parity preserving Pareek Gate. The First Pareek Gate is working as a Master and other one is working as a slave. The design has 3 garbage outputs, 2 constant inputs and 14 quantum cost. The realization of fault tolerant reversible master-slave D flip-flop is shown in Figure 4.5.

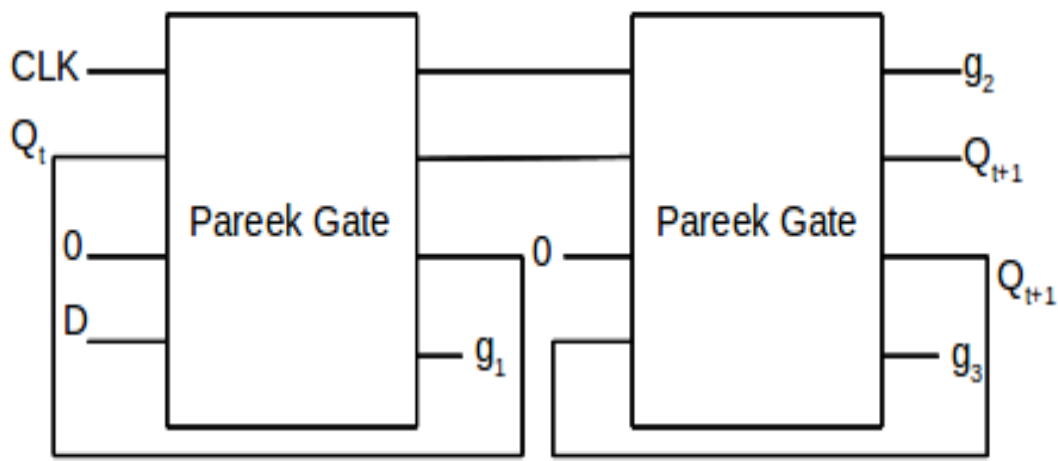

Figure 4.5  Proposed Fault Tolerant Master-Slave D Flip-Flop

### 4.1.4 Fault Tolerant Reversible Double Edge Triggered (DET) D Flip-Flop

The double edge triggered flip-flop is the sequential circuit which samples and stores the input data at both the positive and negative edge of the clock cycles. The frequency of DET flip-flop is reduced to half of the master slave flip-flop. Thus for the low power applications this circuits can be used because frequency is proportional to the power. In the proposed design of fault tolerant reversible DET flip-flop a positive enabled and negative enable D flip-flop are placed in parallel with two Fredkin Gates as shown in Figure 4.6.

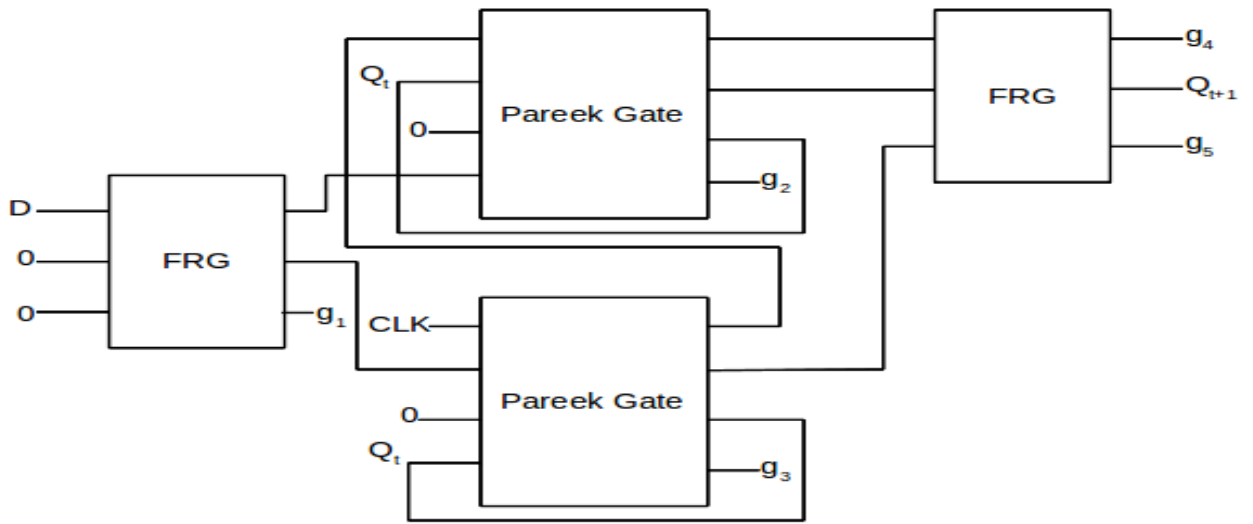

Figure 4.6   Proposed Fault Tolerant Double Edge Triggered (DET) D Flip-Flop

The DET D flip-flop has 4 Gates, 5 garbage outputs, 4 constant inputs and 24 quantum cost. The realization of fault tolerant reversible double edge triggered D flip-flop is shown in Figure 4.6.

### 4.1.5 Fault Tolerant Reversible Positive Level T Flip-Flop

The characteristic equation of the T flip-flop can be written as $Q_{t+1} = (T \oplus Q_t).CLK + CLK'.Q_t$. But the required result can also be obtained from $Q_{t+1} = (T.CLK) \oplus Q_t$. Figure 4.7 shows the proposed design of fault tolerant reversible T flip-flop. The design has one Pareek gate and one F2G. It has 2 constant inputs, 2 garbage outputs and 9 quantum cost.

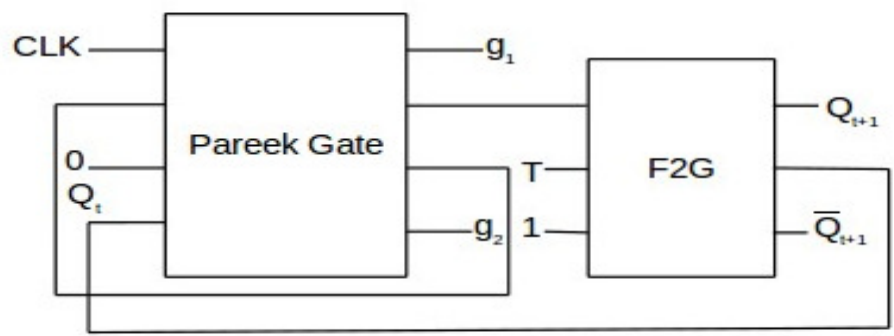

Figure 4.7   Proposed Fault Tolerant Positive Level T Flip-Flop

### 4.1.6 Fault Tolerant Reversible Positive Level Triggered JK Flip-Flop

The characteristic equation of the JK flip-flop can be written as $Q_{t+1} = (JQ_t' + K'Q_t).CLK + CLK'.Q_t$. Figure 4.8 shows the proposed design of fault tolerant reversible JK flip-flop, in which the first Fredkin Gate produces $K'$, which is passed to the second Fredkin Gate to generate $JQ_t$' + $K'Q_t$. The output $JQ_t$' + $K'Q_t$ produced by the second Fredkin Gate is passed to the Pareek Gate and finally to the Double Feynman Gate. The design has one Pareek Gate, two Fredkin and one Double Feynman Gate. It has 5 constant inputs, 6 garbage outputs and 19 quantum cost.

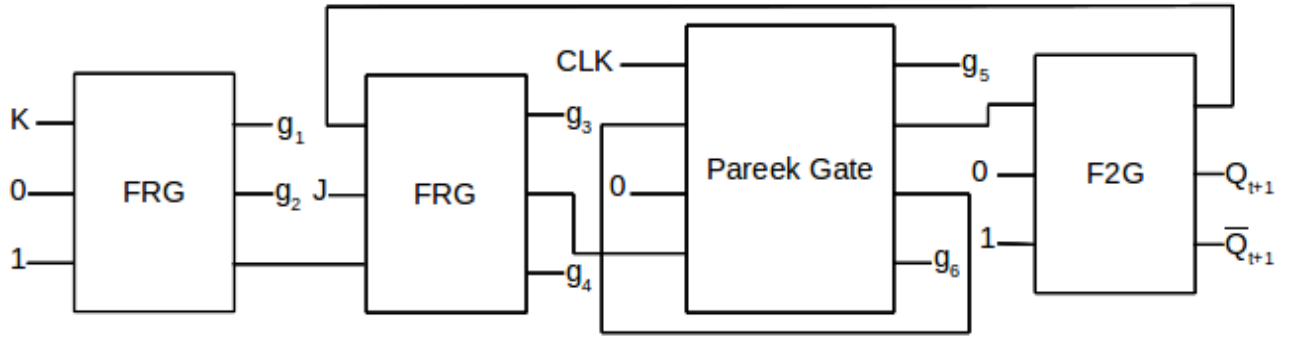

Figure 4.8   Proposed Fault Tolerant Positive Level JK Flip-Flop

### 4.1.7 Fault Tolerant Reversible Positive Level R-S Flip-Flop

The characteristic equation of the R-S flip-flop can be written as $Q_{t+1}$ = (S.CLK + (R.CLK)'$Q_t$). Figure 4.9 shows the proposed design of fault tolerant reversible R-S flip-flop, in which the first Fredkin Gate produces $R'Q_t$, which is passed to the second Fredkin Gate to produce S + $R'Q_t$. The output S + $R'Q_t$ is passed to the Pareek Gate and finally to the Double Feynman Gate to generate the SR flip-flop. The design has one Pareek Gate, two Fredkin and one Double Feynman Gate. It has 5 constant inputs, 6 garbage outputs and 19 quantum cost.

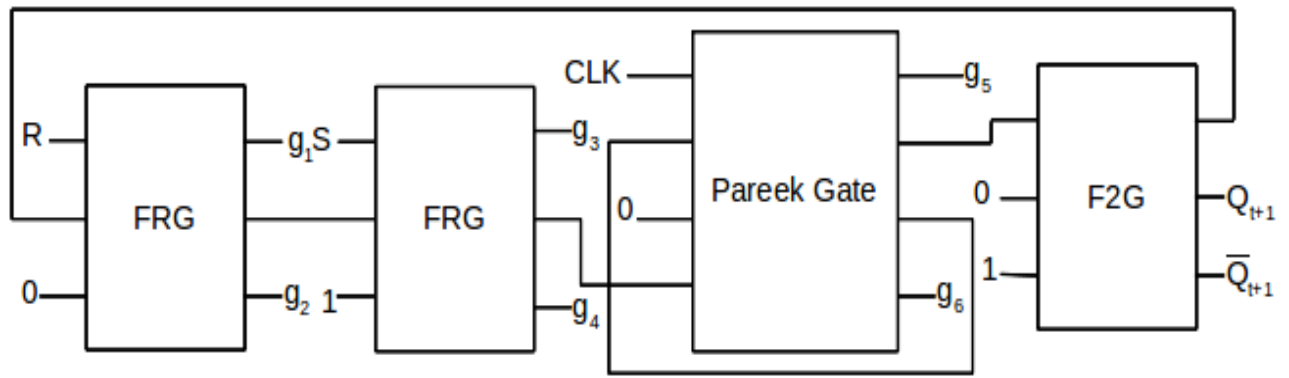

Figure 4.9   Proposed Fault Tolerant Positive Level R-S Flip-Flop

## 4.2  Offline Testable Positive Level D Flip-Flop

The design of offline testable positive level D flip-flop can be tested by only two test sets, all 0s and all 1s, for any universal stuck-at-faults. In this design, we cascade a Fredkin Gate to output $Q_{t+1}$ of the previously proposed positive level D flip flop. Fredkin Gate second input $C_1$ and third input $C_2$ are control input in this design. Stuck-at-fault will be detected with particular test set by breaking the feedback.

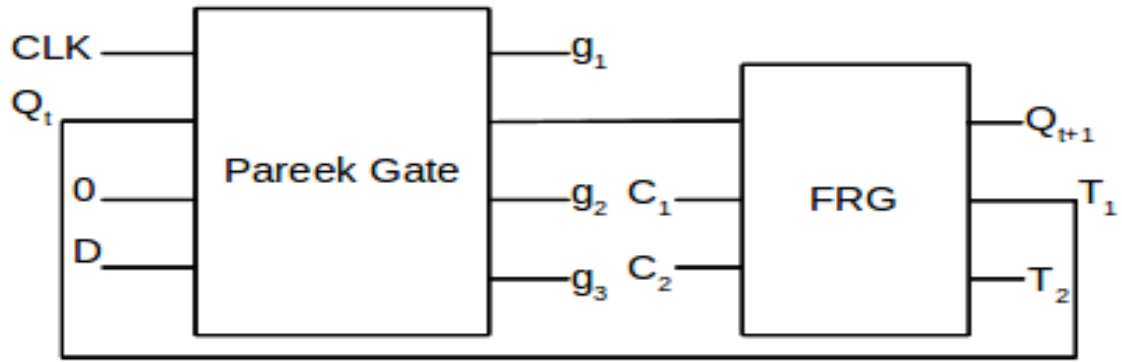

Figure 4.10 Proposed Offline Testable Positive Level D Flip-Flop

For normal execution of positive level D flip-flop value of control inputs are $C_1C_2 = 01$. When $C_1C_2 = 00$, the output $T_1$ will become 0 and the design will become testable with all 0s input vectors for any stuck-at-1 fault. When $C_1C_2 = 11$, the output $T_1$ will become 1 and the design will become testable with all 1s input vectors for any stuck-at-0 fault. Figure 4.10 shows the proposed design of offline testable positive level D flip-flop.

## 4.3 Offline Testable Negative Level D Flip-Flop

The proposed design of offline testable negative level D flip-flop can be tested by only two test sets, all 0s and all 1s, for any universal stuck-at-faults. In this design, we cascade a Fredkin Gate to output $Q_{t+1}$ of the previously proposed negative level D flip-flop. Offline stuck-at-fault detection in negative level D flip-flop is same as the offline testable positive level D flip-flop. Figure 4.11 shows the proposed design of offline testable negative level D flip-flop.

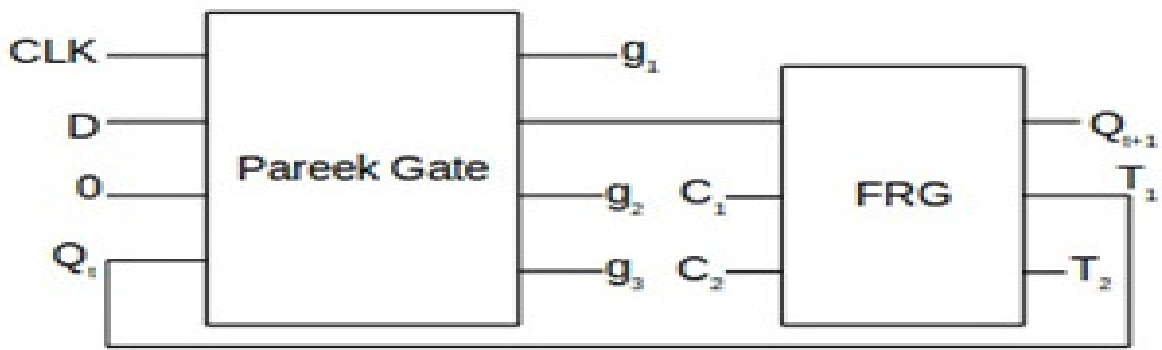

Figure 4.11 Proposed Offline Testable Negative Level D Flip-Flop

## 4.4 Online Testable Positive Level D Flip-Flop

Online testable flip-flop is realized to detect any single bit fault at the output. Figure 4.12 shows the proposed design of online testable positive level D flip-flop. To implement the online testability in previously proposed reversible positive level D flip-flop, we need to add an extra line 'L' with constant input '0'. Afterwards, 3 CNOT Gates are added from second, third and fourth output of the positive level D flip-flop to line L to count parity of these 3 output lines. After computing the 3 CNOT Gates, one more CNOT Gate is added from line L to fourth output of the positive level D flip-flop.

In this online testable circuit, the first input signal to the reversible d flip-flop (clock) will remain same at the output. So, due to parity preserving nature the parity of rest of inputs will be same of the rest of outputs. For online testability of the circuit to detect single bit fault at output, we need to verify the fourth output signal 'T' only. In non-faulty operation (no bit fault at output), the output T will be set 0 due to parity preserving nature of this D flip-flop realization. In faulty operation, the single bit fault occurs at the second or third output signal lines then the fourth testable output T will be set 1. In this situation, the two possibilities of occurrence of single bit fault arise.

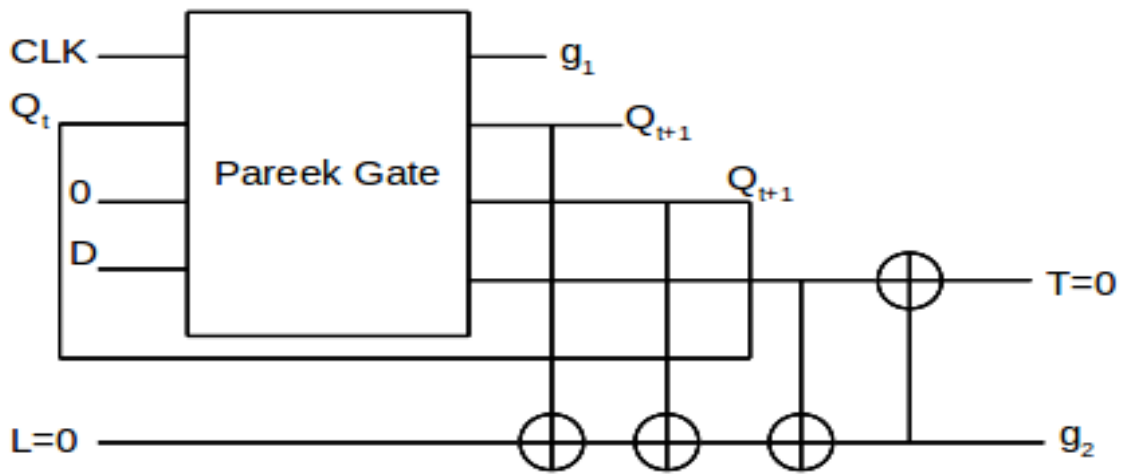

Figure 4.12 Proposed Online Testable Positive Level D Flip-Flop

Case 1: *Bit Fault at Second Output Line $Q_{t+1}$*: In this case, value of second output $Q_{t+1}$ flips 0 to 1 or vice versa.

$L = 0 \oplus Q_{t+1}' \oplus Q_{t+1} \oplus T$

$= T'$                                                           (1)

The X-OR operation is performed between constant input 0 of line L and second output $Q_{t+1}$' (due to bit fault). After that, X-OR operation is performed again between the previous operation output and third output $Q_{t+1}$. Finally, the recent output and fourth output T are computed by X-OR operation to generate the resultant output.

$$T = T \oplus L \text{ (put the value of L from equation (1))}$$
$$= T \oplus T' \qquad\qquad (2)$$
$$= 1$$

The X-OR operation is performed between fourth output T and line L. Now, put the value of line L from equation (1). The equation (2) performs the X-OR operation between fourth output T and value of line L (T' from equation (1)). The resultant output is 1.

Case 2: *Bit Fault at Third Output Line $Q_{t+1}$ Used as Feedback:* In this case, value of third output $Q_{t+1}$ flips 0 to 1 or vice versa.

$$L = 0 \oplus Q_{t+1} \oplus Q_{t+1}' \oplus T$$
$$= T' \qquad\qquad (3)$$

The X-OR operation is performed between constant input 0 of line L and second output $Q_{t+1}$ (due to bit fault). After that, X-OR operation is performed again between the previous operation output and third output $Q_{t+1'}$. Finally, the recent output and fourth output T are computed by X-OR operation to generate the resultant output.

$$T = T \oplus L \text{ (put the value of L from equation (3))}$$
$$= T \oplus T' \qquad\qquad (4)$$
$$= 1$$

The X-OR operation is performed between fourth output T and line L. The equation (4) performs the X-OR operation between fourth output T and value of line L (T' from equation (3)). The resultant output is 1. If the resultant output is 1 in above discussed two cases, then there is an assurance of bit fault at outputs.

In this chapter, we have presented fault tolerant design of sequential building blocks and testable design of D flip-flop. Offline testable design of D flip-flop capable to detect single stuck- at fault and online testable design of D flip-flop detects single bit fault.

# RESULTS

This chapter illustrates statistics and comparison of our design of sequential elements against the proposed designs by various researchers. We use the optimization parameters like gate count, garbage output, constant input, quantum cost and hardware complexity to measure the quality of the design. To compare each sequential elements design a separate table is shown. Each table is separated into five parts as columns by these five optimization parameters. Table rows show our proposed design and existing designs by various researchers. The row for **"Improvement"** is the percentage factor of {100-(Proposed Design/Existing Design)*100} %. For example, for reversible positive D flip-flop design, our realization has 1 gate count while Existing Design [28] has 5 gates. Thus, the ratio is {100-(1/5)*100} % = 80%. The cost of constant input, quantum cost and hardware complexity is not summarized by all the researchers of their reversible flip-flops. We count these numbers based on their designs and show them in tables. To comparison our designs, sections are includes in this chapter as follows:

## 5.1 Statistics and Comparison of Proposed Reversible Building Blocks

In this section proposed design of reversible building blocks such as level triggered D flip-flop, R-S flip-flop, JK flip-flop, T flip-flop, SIPO shift register, PISO shift register and shift register counter are compared with existing designs in literature by comparison tables.

Table 5.1   STATISTICS & COMPARISON OF REVERSIBLE D FLIP-FLOP OVER VARIOUS OPTIMIZATION PARAMETERS

| | Gate Count | Garbage Output | Constant Input | Quantum Cost | Hardware Complexity |
|---|---|---|---|---|---|
| **J. E. Rice [26]** | 11 | 12 | 12 | 47 | $16\alpha+24\beta+5\delta$ |
| **S. K. S. Hari et al. [27]** | 4 | 3 | 2 | 12 | $6\alpha+8\beta+2\delta$ |
| **Lun Chuang et al. [28]** | 5 | 3 | 2 | 13 | $6\alpha+8\beta+3\delta$ |
| **V. Rajmohan et al. [31]** | 1 | 2 | 1 | 8 | $4\alpha+4\beta+\delta$ |
| **Our Design** | *1* | *2* | *1* | *7* | *$3\alpha+2\beta+\delta$* |
| **Improvement w.r.t.[26]** | 91% | 83.3% | 91.6% | 85% | Improved |
| **Improvement w.r.t.[27]** | 75% | 33.3% | 50% | 41.6% | Improved |
| **Improvement w.r.t.[28]** | 80% | 33.3% | 50% | 46% | Improved |
| **Improvement w.r.t.[31]** | 0% | 0% | 0% | 12.5 | Improved |

Table 5.2   STATISTICS & COMPARISON OF REVERSIBLE R-S FLIP-FLOP OVER VARIOUS OPTIMIZATION PARAMETERS

| | Gate Count | Garbage Output | Constant Input | Quantum Cost | Hardware Complexity |
|---|---|---|---|---|---|
| **A. Banerjee et al. [9]** | 9 | 6 | 5 | 33 | $8\alpha+6\beta+\delta$ |
| **H. Thapliyal et al. [25]** | 6 | 8 | 6 | 18 | $10\alpha+12\beta+8\delta$ |
| ***Our Design*** | *4* | *4* | *2* | *13* | *$5\alpha+3\beta+\delta$* |
| **Improvement w.r.t.[9]** | 55% | 33% | 60% | 60% | Improved |
| **Improvement w.r.t.[25]** | 33% | 50% | 66% | 27% | Improved |

Table 5.3   STATISTICS & COMPARISON OF REVERSIBLE JK FLIP-FLOP OVER VARIOUS OPTIMIZATION PARAMETERS

| | Gate Count | Garbage Output | Constant Input | Quantum Cost | Hardware Complexity |
|---|---|---|---|---|---|
| **J. E. Rice [26]** | 12 | 14 | 13 | 52 | $17\alpha+26\beta+5\delta$ |
| **Lun Chuang et al. [28]** | 7 | 4 | 3 | 27 | $7\alpha+9\beta+2\delta$ |
| ***Our Design*** | *4* | *4* | *2* | *13* | *$6\alpha+6\beta+3\delta$* |
| **Improvement w.r.t.[26]** | 66% | 71% | 84% | 75% | Improved |
| **Improvement w.r.t.[28]** | 42% | 0% | 33% | 51% | Improved |

Table 5.4   STATISTICS & COMPARISON OF REVERSIBLE T FLIP-FLOP OVER VARIOUS OPTIMIZATION PARAMETERS

| | Gate Count | Garbage Output | Constant Input | Quantum Cost | Hardware Complexity |
|---|---|---|---|---|---|
| **J. E. Rice [26]** | 13 | 14 | 14 | 53 | $18\alpha+26\beta+5\delta$ |
| **S. K. S. Hari et al. [27]** | 5 | 3 | 2 | 13 | $7\alpha+8\beta+2\delta$ |
| **A. Banerjee et al. [36]** | 3 | 3 | 2 | 17 | $9\alpha+8\beta+2\delta$ |
| ***Our Design*** | *2* | *2* | *1* | *8* | *$4\alpha+2\beta+\delta$* |
| **Improvement w.r.t.[26]** | 86% | 85% | 93% | 85% | Improved |
| **Improvement w.r.t.[27]** | 60% | 33% | 50% | 39% | Improved |
| **Improvement w.r.t.[36]** | 33% | 33% | 50% | 53% | Improved |

Table 5.5   STATISTICS & COMPARISON OF REVERSIBLE SIPO SHIFT REGISTER OVER VARIOUS OPTIMIZATION PARAMETERS

| | Gate Count | Garbage Output | Constant Input | Quantum Cost | Hardware Complexity |
|---|---|---|---|---|---|
| **V. Rajmohan et al. [31]** | 7 | 5 | 7 | 35 | $19\alpha+16\beta+4\delta$ |
| **Our Design** | *7* | *5* | *7* | *31* | *$15\alpha+8\beta+4\delta$* |
| **Improvement w.r.t.[31]** | 0% | 0% | 0% | 11% | Improved |

Table 5.6   STATISTICS & COMPARISON OF REVERSIBLE PISO SHIFT REGISTER OVER VARIOUS OPTIMIZATION PARAMETERS

| | Gate Count | Garbage Output | Constant Input | Quantum Cost | Hardware Complexity |
|---|---|---|---|---|---|
| **V. Rajmohan et al. [31]** | 8 | 10 | 4 | 52 | $23\alpha+32\beta+8\delta$ |
| **Our Design** | *8* | *10* | *4* | *48* | *$24\alpha+24\beta+8\delta$* |
| **Improvement w.r.t.[31]** | 0% | 0% | 0% | 11% | Improved |

Table 5.7   STATISTICS OF REVERSIBLE SHIFT REGISTER COUNTER

| | Gate Count | Garbage Output | Constant Input | Quantum Cost | Hardware Complexity |
|---|---|---|---|---|---|
| **Proposed Design** | *8* | *5* | *8* | *32* | *$16\alpha+8\beta+4\delta$* |

## 5.2 Statistics & Comparison of Proposed Fault Tolerant Reversible Blocks

In this section proposed design of fault tolerant reversible building blocks such as positive level triggered D flip-flop, negative level Triggered D flip-flop, master- slave D flip-flop, double edge triggered D flip-flop, level triggered R-S flip-flop, level triggered JK flip-flop and level triggered T flip-flop are compared with existing designs in literature by comparison tables. States for offline testable designs are also shown in tables.

Table 5.8  STATISTICS & COMPARISON OF REVERSIBLE FAULT TOLERANT POSITIVE LEVEL D FLIP-FLOP OVER VARIOUS OPTIMIZATION PARAMETERS

| | Gate Count | Garbage Output | Constant Input | Quantum Cost | Hardware Complexity |
|---|---|---|---|---|---|
| **H. Thapliyal et al. [34]** | 2 | 3 | 2 | 10 | *4α+8β+2δ* |
| **A. Hatam et al. [20]** | 2 | 2 | 1 | 7 | *4α+4β+δ* |
| **Our Design** | *1* | *2* | *1* | *7* | *3α+2β+δ* |
| **Improvement w.r.t.[34]** | 50% | 33% | 50% | 30% | Improved |
| **Improvement w.r.t.[20]** | 50% | 0% | 0% | 0% | Improved |

Table 5.9  STATISTICS & COMPARISON OF REVERSIBLE FAULT TOLERANT NEGATIVE LEVEL D FLIP-FLOP OVER VARIOUS OPTIMIZATION PARAMETERS

| | Gate Count | Garbage Output | Constant Input | Quantum Cost | Hardware Complexity |
|---|---|---|---|---|---|
| **H. Thapliyal et al. [34]** | 2 | 3 | 2 | 10 | *4α+8β+2δ* |
| **Our Design** | *1* | *2* | *1* | *7* | *3α+2β+δ* |
| **Improvement w.r.t.[34]** | 50% | 33% | 50% | 30% | Improved |

Table 5.10  STATISTICS & COMPARISON OF REVERSIBLE FAULT TOLERANT MASTER-SLAVE D FLIP-FLOP OVER VARIOUS OPTIMIZATION PARAMETER

| | Gate Count | Garbage Output | Constant Input | Quantum Cost | Hardware Complexity |
|---|---|---|---|---|---|
| **H. Thapliyal et al. [34]** | 4 | 5 | 4 | 20 | *8α+16β+4δ* |
| **A. Hatam et al. [20]** | 3 | 4 | 3 | 16 | *10α+8β+2δ* |
| **Our Design** | *2* | *3* | *2* | *14* | *6α+4β+2δ* |
| **Improvement w.r.t.[34]** | 50% | 40% | 50% | 30% | Improved |
| **Improvement w.r.t.[20]** | 33% | 25% | 33% | 13% | Improved |

Table 5.11  STATISTICS OF REVERSIBLE FAULT TOLERANT T FLIP-FLOP

| | Gate Count | Garbage Output | Constant Input | Quantum Cost | Hardware Complexity |
|---|---|---|---|---|---|
| **Our Design** | *2* | *2* | *2* | *9* | *5α+2β+δ* |

Table 5.12  Statistics & Comparison of Reversible Fault Tolerant DET D Flip-Flop Over Various Optimization Parameter

| | Gate Count | Garbage Output | Constant Input | Quantum Cost | Hardware Complexity |
|---|---|---|---|---|---|
| H. Thapliyal et al. [34] | 6 | 7 | 6 | 30 | *12α+24β+6δ* |
| Our Design | *4* | *5* | *4* | *24* | *10α+12β+4δ* |
| Improvement w.r.t.[34] | 33% | 29% | 33% | 20% | Improved |

Table 5.13  Statistics of Reversible Fault Tolerant JK Flip-Flop

| | Gate Count | Garbage Output | Constant Input | Quantum Cost | Hardware Complexity |
|---|---|---|---|---|---|
| Our Design | *4* | *6* | *5* | *19* | *9α+10β+3δ* |

Table 5.14  Statistics of Reversible Fault Tolerant R-S Flip-Flop

| | Gate Count | Garbage Output | Constant Input | Quantum Cost | Hardware Complexity |
|---|---|---|---|---|---|
| Our Design | *4* | *6* | *5* | *19* | *9α+10β+3δ* |

Table 5.15  Statistics of Reversible Offline Testable Positive D Flip-Flop

| | Gate Count | Garbage Output | Constant Input | Quantum Cost | Hardware Complexity |
|---|---|---|---|---|---|
| Our Design | *2* | *3* | *3* | *12* | *5α+6β+2δ* |

The above statistics clearly show that improvement in designs of sequential elements compare to existing design.

# CONCLUSION AND FUTURE SCOPE

This dissertation work began by placing reversible computation in its proper historical context - starting with the issues related to sequential reversible circuit design and continuing to the incorporation of fault tolerance capability in reversible sequential circuit. From this foundation, the rest of the dissertation focused on the optimized designs of reversible flip-flop which are the most significant and basic memory elements that will be the target building block of memory for the next generation computing devices, optimized design of sequential circuits and finally fault tolerance and testable designs of reversible flip-flop. The following sections recount the major contributions in each of these aspects.

## 6.1 Contribution to Design of Reversible Sequential Building Blocks

In this dissertation, we have proposed a novel parity preserving Pareek Gate and a complete set of reversible sequential elements corresponding to available irreversible sequential designs. Our proposed reversible flip–flops realization are significantly improved over existing reversible realization in terms of gate count, garbage output, constant input, and quantum cost and hardware complexity. We have also proposed a quantum realization of our proposed gate. This is a significant contribution, for instance, for the design of reversible sequential circuits using emerging devices.

## 6.2 Contribution to Design of Reversible Sequential Circuits Design

This dissertation also presents novel designs of shift registers and shift register counter. These proposed reversible realizations are significantly improved over existing reversible realization in terms of gate count, garbage output, constant input, and quantum cost and hardware complexity.

## 6.3 Contribution to Fault Tolerant &Testable Design of Sequential Blocks

This dissertation also presented novel designs of fault tolerant reversible sequential elements such as positive and negative level flip-flops, master-slave flip-flop and double edge triggered flip-flop. Afterwards, this dissertation proposes the offline testable approach for positive and negative level D flip-flop. The online testable design of positive level D flip-flop has been proposed for the first time. The design process is also introduced in detail. Our designs are very cost effective compare to exiting designs. The importance of this dissertation lies in the fact that it provides the offline and online testable design for positive and negative level D flip-flop completely testable for the stuck-at fault and single bit fault at output in the circuit respectively. This work forms an important step in building of fault tolerant sequential reversible circuits and fault detection in reversible computing.

## 6.4 Future Scope

Though the foundation for reversible computation were laid more than fifty years ago, and interest of researcher have increased considerably in last decade, in many ways the area is still fresh. In this dissertation, some of the unknown aspects of reversible computing were explored, but there is still plenty to explore for the future. There are clear and abundant areas for future work stemming from this dissertation work. Some of them are listed below:

➢ Building such an optimized design of reversible complex sequential circuits taking advantage of the proposed optimized design of reversible building blocks comprises the future work.

➢ This dissertation proposes fault tolerance design through parity-preserving reversible logic design which provides detection of single-bit fault occurred in the circuit. The conservative design providing multiple fault detection is still unexplored. Further Investigations are required in the field of multiple-fault detection by conservative logic design for reversible sequential computing.

➢ This dissertation work can be extended in detecting the fault in more complex sequential circuit like counters, registers. Multiple bit fault detection and its correction through cyclic redundancy check (CRC) circuit design can be the future work in the field of sequential reversible circuits.

# PAPER PUBLICATION OUT OF DISSERTATION